\documentclass[reprint,prb,amsmath,amssymb,aps,showpacs,preprintnumbers]{revtex4-2}
\usepackage{graphicx} % Include figure files
\usepackage{datetime}
\usepackage{bm} % bold math

\begin{document}
\title{Non-Markovian effects for hybrid plasmonic systems in the strong coupling regime}
\author{Tigran V. Shahbazyan}
\affiliation{Department of Physics, Jackson State University, Jackson, Mississippi 39217 USA}

%\date{\today,\,\,\currenttime}

\begin{abstract} 
We study the role of non-Markovian effects in the emission spectrum of a quantum emitter resonantly coupled to a surface plasmon in a metal-dielectric structure as the system transitions to strong coupling regime. By using a recent quantum approach to interacting plasmons that incorporates the effects of host material's optical dispersion and losses in the coupling parameters, we obtain analytically the emission spectrum for a plasmonic system of arbitrary shape with characteristic size below the diffraction limit. In the weak coupling regime, the dispersion-induced non-Markovian  effects are weak and do not significantly affect the spectral shape of the emission peak. In contrast, in the strong coupling regime, the non-Markovian effects lead to dramatic changes in the emission spectra by causing inversion of spectral asymmetry, as compared with classical and quantum models based on the Markov approximation, which results in a strong enhancement of the lower frequency polaritonic band, consistent with the experiment.
\end{abstract}
\maketitle

%%%%%%%%%%%%%%%%%%%%%%%%%%%%%%%%%%%%%%%%%%%%%%%
\section{Introduction}

The effects of strong coupling between surface plasmons in metal-dielectric structures  and excitons in semiconductors or dye molecules have recently attracted considerable interest driven  by numerous potential applications including ultrafast reversible switching \cite{ebbesen-prl11,bachelot-nl13,zheng-nl16}, quantum computing \cite{waks-nnano16,senellart-nnano17} or light harvesting \cite{leggett-nl16}. In the strong coupling regime, coherent energy exchange between excitons and plasmons  \cite{shahbazyan-nl19,mortensen-rpp20} leads to the emergence of mixed polaritonic states with energy bands separated by the anticrossing gap (Rabi splitting) \cite{novotny-book}. While Rabi splittings in the emission spectra of excitons coupled to  cavity modes in semiconductor microcavities are about several meV  \cite{forchel-nature04,khitrova-nphys06,imamoglu-nature06}, they can reach hundreds  meV in hybrid plasmonic systems involving excitons in \textit{J}-aggregates \cite{bellessa-prl04,sugawara-prl06,wurtz-nl07,fofang-nl08,bellessa-prb09,schlather-nl13,lienau-acsnano14,shegai-prl15}, in various dye molecules \cite{hakala-prl09,berrier-acsnano11,salomon-prl12,luca-apl14,noginov-oe16}  or in semiconductor nanostructures \cite{vasa-prl08,gomez-nl10,gomez-jpcb13,manjavacas-nl11}  resonantly coupled to surface plasmons. For single excitons, however, reaching  a strong coupling regime is a  challenging task as it requires extremely small plasmon mode volumes that can mainly be achieved in nanogaps \cite{hecht-sci-adv19,pelton-sci-adv19,baumberg-natmat2019}.

At the same time, the shape of emission spectra in the strong coupling regime  remains an actively debated issue as the relative spectral weight of polaritonic bands is determined by several competing processes.
%\cite{savvidis-aom13,ebbesen-fd15,ebbesen-nc15,shegai-nl17,shegai-acsphot19,zhang-nl17,xu-nl17,hughes-optica15,garsia-vidal-njp15,ding-prl17,aizpurua-optica18,shahbazyan-nanophot21}. 
In general, the emission spectrum of a hybrid system, characterized by (effective) dipole moment $\bm{p}_{s}$, is $\propto|\bm{p}_{s}|^{2}\omega^{4}$ \cite{novotny-book}, where  $\omega$ is the emission frequency, implying that, in the strong coupling regime, the spectral band associated with upper-energy polaritonic state should be relatively enhanced due to its higher radiation rate. Such a spectral profile is, in fact, predicted by the widely-used classical model of two coupled oscillators (CO), in which only one oscillator (plasmon) couples to the electromagnetic (EM) field while the second (exciton) is treated as dark due to its much smaller optical dipole moment \cite{pelton-oe10,pelton-nc18,pelton-ns19}.  However, recent experiments for excitons resonantly coupled to cavity modes in semiconductor microcavities \cite{savvidis-aom13,ebbesen-fd15,ebbesen-nc15} or surface plasmons in metal-dielectric structures \cite{shegai-nl17,shegai-acsphot19,zhang-nl17,xu-nl17} reveal the \textit{opposite} asymmetry pattern characterized by enhanced \textit{lower} polaritonic band. For plasmonic systems, a shift of spectral weight  in the absorption and scattering spectra has been obtained by including the Fano interference effects between the plasmon's and plasmon-induced exciton's dipole moments; however, a significant spectral weight shift would require an extremely strong field confinement \cite{ding-prl17,shahbazyan-nanophot21,xu-acsphot21} or a large number of excitons coupled to the plasmon \cite{shahbazyan-jcp22}. At the same time, for molecular excitons coupled to a cavity mode, in the absence of strong field confinement in microcavities, the accurate spectral weight of polaritonic bands in the emission spectra is obtained, within quantum master equation approach, by incorporating excitations of vibronic modes that  accompany optical transitions \cite{garsia-vidal-njp15, aizpurua-optica18,settineri-pra18}.
% Using the above approach, it has been shown that exciton interactions with vibronic modes strongly affect the spectral shape of surface-enhanced Raman scattering \cite{hughes-prb21}.

At the same time, the effects of optical dispersion and losses in  metal-dielectric structures,  characterized by a frequency-dependent complex dielectric function $\varepsilon (\omega,\bm{r})=\varepsilon'(\omega,\bm{r})+i\varepsilon''(\omega,\bm{r})$, are far more significant than for semiconductor microcavities. Here we stress that the non-Markovian effect in plasmonic systems are distinct from those emerging from the interactions of quantum emitters (QEs) with the reservoir of photon or phonon states  \cite{nori-prb09,averkiev-jetp09,tejedor-prb10,thanopulos-prb17,moradi-sr18,molmer-acsph19}. In the latter cases, the QE interactions with the reservoir states give rise to memory effects, which lead to nonexponential time evolution of the emission signal and to its oscillations with characteristic time depending on the QE-reservoir coupling. In contrast, surface plasmons at metal-dielectric interfaces interact \textit{directly} with the EM field via their own \textit{frequency-dependent} optical dipole moment $\bm{\mu}(\omega)$ (see below) that is much larger than the QE dipole moment. Furthermore, in the \textit{strong coupling regime}, the QE-plasmon coupling is comparable to the overall plasmon decay rate, which, e.g., for gold structures, corresponds to $\approx 10$ fs lifetime \cite{stockman-review}, implying that, close to the transition point, observing non-Markovian effects for plasmonic systems by means of time-resolved spectroscopy represents a formidable challenge. However, as we show in this paper, the dispersion-induced non-Markovian effects can show up prominently in the optical spectra of plasmonic systems due to a rather nontrivial interplay between the metal dielectric function's real part $\varepsilon'(\omega,\bm{r})$ and its  imaginary part $\varepsilon''(\omega,\bm{r})$, which define, respectively, the plasmon coupling to the EM field and the broad plasmon optical band. Specifically, the strong frequency dispersion of $\varepsilon'(\omega,\bm{r})$, which originates from the free-electron absorption is metals, can lead to a prominent spectral weight redistribution between the lower and higher frequency parts of the broad emission spectra. In the strong coupling regime characterized by well-separated polaritonic bands, such non-Markovian effects result, in fact, in an \textit{inversion of spectral asymmetry} as compared with the Markovian calculations.

Note that the materials' optical dispersion effects in optical spectra cannot be described neither within the classical CO model with phenomenological parameters nor within the master equation approaches based on the canonical Hamiltonian for plasmon modes as the latter is   only valid in the Markov approximation \cite{shahbazyan-prb21}. In principle, the dispersion effects can be accounted for within the macroscopic quantum electrodynamics approach based on the fluctuation-dissipation theorem  which  involves  the reservoir states of the entire metal-dielectric structure \cite{welsch-pra98,welsch-p00,philbin-njp10}. However, the excessively large Hilbert space of reservoir states that are extended over the entire system volume makes it difficult,  except for relatively simple systems \cite{welsch-pra00,dzsotjan-prb10,andreani-prb12,hughes-prb12,hughes-prb13,zubairy-prb14,garcia-prl14,rousseaux-prb18,sivan-prb19}, to model quantum dynamics  of plasmons excited primarily at the metal-dielectric interfaces. Another numerical approach to modeling of electromagnetic excitations in metal-dielectric structures involves expansion over a set of quasinormal modes \cite{hughes-njp14,hughes-acsphot14,lalanne-pra14,hughes-pra15,muljarov-prb16,lalanne-prb18,lalanne-lpr18}.  

In this paper, to elucidate the emergence of non-Markovian effects in optical spectra of hybrid plasmonic systems, we restrict ourselves to relatively small plasmonic systems with characteristic size below the diffraction limit. To this end, we employ a novel analytical quantum approach developed in our recent paper \cite{shahbazyan-prb21} that treats localized plasmons as electronic excitations interacting with QEs and the EM field. In this approach, starting with macroscopic electrodynamics quantization scheme \cite{welsch-pra98,welsch-p00,philbin-njp10}, the system's reservoir states are projected upon localized plasmon modes, thereby reducing the full Hilbert space to a much more limited subspace spanned by a discrete set of bosonic operators with linear dispersion. Interactions of these projected reservoir modes (PRMs) with QEs and the EM field are mediated by  \textit{classical} plasmons, while the   coupling parameters are defined explicitly by the plasmon local fields, system geometry and  frequency-dependent dielectric function of the host material \cite{shahbazyan-prb21}. Using this approach, we address the role of dispersion-induced non-Markovian effects in the emission spectrum of a QE resonantly coupled to a localized surface plasmon in a metal-dielectric structure as this system transitions to strong coupling regime. We show that, in the weak coupling regime, the non-Markovian effects  are weak and do not significantly affect the emission spectrum shape. However, as the system transitions to strong coupling regime characterized by distinct polaritonic  bands, the non-Markovian effects dramatically affect the emission spectrum by shifting the spectral weight towards the lower polaritonic band, in striking contrast with Markov-approximation-based calculations. For a QE near a plasmonic nanostructure of\textit{ arbitrary} shape but with the characteristic size below the diffraction limit, we  obtain an explicit expression for the radiated power spectrum in terms of frequency-dependent coupling parameters and elucidate the relevant processes contributing to the emission. In the strong coupling regime, we perform numerical calculations for a QE situated near the tip of a gold nanorod illustrating inversion of spectral asymmetry caused by dispersion-induced non-Markovian effects.
%reported in the experiment \cite{asymmetry}.

The paper is organized as follows. In Sec. \ref{sec:quantum} we outline  our approach to interacting quantum plasmon and collect the relevant formulas to be used in the following sections. In Sec. \ref{sec:emission}, we derive the emission power spectrum for a QE resonantly coupled to a localized plasmon mode in a metal-dielectric structure. In Sec. \ref{sec:weak}, we elucidate the processes contributing to the emission spectrum in the weak coupling regime.  In Sec. \ref{sec:strong}, we present our numerical results for the emission spectrum of a QE near a Au nanorod as this hybrid system transitions to the strong coupling regime. Section \ref{sec:conc} concludes the paper. In the appendix,  some technical detail  of our  approach are elucidated.

\section{Quantum approach to interacting plasmons in metal-dielectric structures}
\label{sec:quantum}

In this section, we outline our recent approach to interacting quantum plasmons in terms of discrete set of bosonic modes with linear dispersion \cite{shahbazyan-prb21} (see also the Appendix). We consider $N$ QEs with excitation frequency  $\omega_{e}$ and dipole moments $\bm{\mu}_{i}=\mu_{e}\bm{n}_{i}$  ($\bm{n}_{i}$ is the dipole orientation of $i$th QE) situated at positions $\bm{r}_{i}$ ($i=1,\dots,N$) near a metal-dielectric structure characterized by a complex dielectric function $\varepsilon (\omega,\bm{r})=\varepsilon' (\omega,\bm{r})+i\varepsilon'' (\omega,\bm{r})$. We assume that the characteristic system size is much smaller than the radiation wavelength, and so the structure supports localized plasmon modes  described by  quasistatic Gauss's equation $\bm{\nabla}\cdot\left [\varepsilon' (\omega_{m},\bm{r})\bm{\nabla} \Phi_{m}(\bm{r})\right ]=0$, where $\omega_{m}$ is the plasmon mode frequency and $\Phi_{m}(\bm{r})$ is the mode potential that defines the mode electric field as  $\bm{E}_{m}(\bm{r})=-\bm{\nabla} \Phi_{m}(\bm{r})$,  chosen to be real here \cite{stockman-review}. Since the classical plasmon mode energy is
\begin{align}
\label{mode-energy}
U_{m}
= \frac{1}{16\pi} 
\!\int \!  dV     
\dfrac{\partial[\omega_{m}\varepsilon'(\omega_{m},\bm{r})]}{\partial \omega_{m}}
\, \bm{E}_{m}^{2}(\bm{r}),
% \nonumber\\
%&= \frac{1}{16\pi} 
%\!\int \!  dV     
%\dfrac{\partial[\omega_{m}\varepsilon'(\omega_{m},\bm{r})]}{\partial \omega_{m}}
% \bm{E}_{m}^{2}(\bm{r}) 
%%\partial [\omega_{m}\varepsilon'(\omega_{m},\bm{r})]/\partial \omega_{m}.
\end{align}
it is convenient to use instead the normalized  fields $\tilde{\bm{E}}_{m}(\bm{r})=\sqrt{\hbar\omega_{m}/4U_{m}} \bm{E}_{m}(\bm{r})$ to describe eigenmodes of quantum plasmons with energy $\hbar\omega_{m}$ \cite{shahbazyan-prb21}. The Hamiltonian of a hybrid QE-plasmon  system interacting with the EM field is
\begin{equation}
\label{H-full}
\hat{H}=\hat{H}_{\rm b}+\hat{H}_{\rm b-qe}+\hat{H}_{\rm b-em}+\hat{H}_{\rm qe}+\hat{H}_{\rm qe-em}.
\end{equation}
Here, the first term,
\begin{equation}
\label{H-b}
\hat{H}_{\rm b}=\sum_{m}  \!\int_{0}^{\infty}\!   d\omega \,\hbar\omega \,\hat{b}^{\dagger}_{m}(\omega)\hat{b}_{m}(\omega), 
 \end{equation} 
is the Hamiltonian for reservoir states projected upon the plasmon modes, to be referred to as projected reservoir modes (PRM), which are described by creation and annihilation operators $\hat{b}^{\dagger}_{m}$ and $\hat{b}_{m}$, respectively, obeying commutation relations $[\hat{b}_{m}(\omega),\hat{b}_{n}^{\dagger}(\omega')]=\delta_{mn}\delta(\omega-\omega')$. 

The second term in the Hamiltonian (\ref{H-full}) describes PRM interactions with QEs and has the form 
\begin{equation}
\label{H-b-qe-main}
\hat{H}_{\rm b-qe}=\sum_{im} \int_{0}^{\infty}\!  d\omega\left [\hbar q_{im}(\omega)\,\hat{\sigma}^{\dagger}_{i}\,\hat{b}_{m}(\omega) +\text{H.c.}
%g_{im}^{*}(\omega)\hat{b}^{\dagger}_{m}(\omega)\hat{\sigma}_{i}
\right ],
\end{equation}
where $\hat{\sigma}^{\dagger}_{i} $ and $\hat{\sigma}_{i} $ are, respectively, the raising and lowering operators for the $i$th QE and $q_{im}(\omega)=g_{im}\lambda_{m}(\omega)$ is QE-PRM coupling. Here,  $g_{im}=-\bm{\mu}_{i}\!\cdot\!\tilde{\bm{E}}_{m}(\bm{r}_{i})/\hbar$ is the standard  QE-plasmon coupling, while the function $\lambda_{m}(\omega)$ is given by \cite{shahbazyan-prb21} (see Appendix)
\begin{align}
\label{lambda-main}
\lambda_{m}(\omega)
= \sqrt{\frac{\gamma_{m}(\omega)}{2\pi}}\frac{ i }{\omega -\omega_{m}+\frac{i}{2}\gamma_{m}(\omega)},
\end{align}
where $\gamma_{m}(\omega)$ is  frequency-dependent nonradiative decay rate of a plasmon mode; in structures with  a single metallic component, it has the standard form \cite{stockman-review} $\gamma_{m}(\omega)=2\varepsilon''(\omega)/[\partial\varepsilon'(\omega_{m})/\partial \omega_{m}]$. Importantly, the function  $\lambda_{m}(\omega)$ has a \textit{plasmon pole} in the complex-frequency plane, implying that QE-PRM interactions are mediated by the classical plasmons \cite{shahbazyan-prb21}. Specifically, the rate of energy transfer (ET) from an excited QE to a plasmonic mode can be obtained in the first order as
\begin{equation}
\label{rate-abs-main}
\gamma_{i\rightarrow m}(\omega)=\frac{2\pi}{\hbar}
\int_{0}^{\infty}\!  d\omega' 
\left | \hbar q_{im}(\omega')\right |^{2}\delta(\hbar\omega'-\hbar\omega),
\end{equation}
where the  integral runs over  PRM's final states, yielding
\begin{equation}
\label{rate-et-mode-main}
\gamma_{i\rightarrow m} (\omega) 
%= 2\pi g_{im}^{2}\left | \lambda_{m}(\omega) \right |^{2}
=\frac{g_{im}^{2} \gamma_{m}(\omega)}{(\omega_{m}-\omega)^{2} +\frac{1}{4}\gamma_{m}^{2}(\omega)}. 
\end{equation}
At resonance  $\omega=\omega_{m}$, we recover the relation between QE-plasmon coupling and QE-plasmon ET rate  as \cite{shahbazyan-nl19} $g_{im}^{2}=\gamma_{i\rightarrow m}\gamma_{m}/4$.
%Thus, the classical effect of resonance ET from a QE to plasmons emerges from the Hamiltonian (\ref{H-b-qe}) in the lowest order. 

The third term  in the Hamiltonian (\ref{H-full}) describes PRM interactions with the EM field. For a monochromatic field of frequency $\omega$  and amplitude $\bm{E}$ uniform on the system scale, in the rotating wave approximation (RWA), this Hamiltonian term has the form 
\begin{equation}
\label{H-b-em-main}
\hat{H}_{\rm b-em}= -\!\sum_{m}\! \int_{0}^{\infty}\! \! d\omega' \! \left [\bm{d}_{m}^{*}(\omega,\omega')\!\cdot \!\bm{E}e^{-i\omega t} \hat{b}_{m}^{\dagger}(\omega') +\text{H.c.}
%g_{im}^{*}(\omega)\hat{b}^{\dagger}_{m}(\omega)\hat{\sigma}_{i}
\right ],
\end{equation}
where $\bm{d}_{m}(\omega,\omega')=\bm{\mu}_{m}(\omega)\lambda_{m}(\omega')$ is optical transition matrix element (star denotes complex conjugate). Here, $\bm{\mu}_{m}(\omega)=\!\int\! dV\chi' (\omega,\bm{r})\tilde{\bm{E}}_{m}(\bm{r})$ is optical dipole moment of the plasmon mode, defined by plasmonic structure's susceptibility $\chi(\omega,\bm{r})=[\varepsilon(\omega,\bm{r})-1]/4\pi$, which determines the frequency-dependent plasmon  radiative decay rate as 
\begin{equation}
\label{mode-decay-rad-main}
\gamma_{m}^{rad}(\omega)=\frac{4\mu_{m}^{2}(\omega)\omega^{3}}{3\hbar c^{3}},
\end{equation}
where $c$ is the speed of light \cite{shahbazyan-prb18}. Note that, similar to  QE-PRM coupling, the factor $\lambda_{m}(\omega')$  implies that PRM interactions with the EM field are mediated by classical plasmons, i.e., for the EM field frequency  close to a plasmon mode frequency, the corresponding optical transition is resonantly enhanced.

The last two terms in the Hamiltonian (\ref{H-full}) are the standard QE Hamiltonian 
\begin{equation}
\label{H-qe-main}
H_{\rm qe}=\hbar\omega_{e}\!\sum_{i}\hat{\sigma}_{i}^{\dagger}\hat{\sigma}_{i},
\end{equation}
and the Hamiltonian term describing  QEs' interactions with the EM field,
\begin{equation}
\label{H-qe-em-main}
H_{\rm qe-em}=-\sum_{i}(\bm{\mu}_{i}\!\cdot\!\bm{E}e^{-i\omega t} \hat{\sigma}_{i}^{\dagger} + {\rm H.c.}).
\end{equation}
We stress that the coupling parameters, characterizing the  interactions of PRMs with QEs and the EM field, retain frequency dependence of the material dielectric function including both its real part [via $\bm{\mu}_{m}(\omega)$] and imaginary part [via $\gamma_{m}(\omega)$]. In contrast, within the canonical quantization scheme, these  parameters are taken at the plasmon  frequency, i.e., $\omega=\omega_{m}$ (Markov approximation) \cite{shahbazyan-prb21}. Note that, in either approach, the QE-plasmon coupling $g_{im}$ is frequency-independent and can be recast,  in terms of the original mode fields, in a cavity-like form,
\begin{equation}
\label{coupling-mode-volume-main}
g^{2}_{im}
%=\frac{4\pi \omega_{m}\mu^{2}[\bm{n}_{i}\!\cdot\!\bm{E}_{m}(\bm{r}_{i})]^{2}}{\hbar\int \! dV [\partial (\omega_{m}\varepsilon')/\partial \omega_{m}]\bm{E}^{2}_{m}}
=\frac{2\pi \mu_{e}^{2}\omega_{m}}{\hbar{\cal V}^{(i)}_{m}},
~~
\frac{1}{{\cal V}^{(i)}_{m}}
%=\frac{[\bm{n}\!\cdot\!\bm{E}(\bm{r})]^{2}}{8\pi U }
= \frac{2[\bm{n}_{i}\!\cdot\!\bm{E}_{m}(\bm{r}_{i})]^{2}}{\int \! dV [\partial (\omega_{m}\varepsilon')/\partial \omega_{m}]\bm{E}_{m}^{2}},
\end{equation}
where ${\cal V}^{(i)}_{m}$ is  the projected  plasmon mode volume characterizing  the plasmon field confinement at a point $\bm{r}_{i}$ in the direction $\bm{n}_{i}$ \cite{shahbazyan-acsphot17,shahbazyan-prb18,shahbazyan-nl19}. In the following sections, we demonstrate that non-Markovian effects due to dispersion of the above parameters have profound effect on the emission spectra in the strong coupling regime.

\section{Spontaneous emission near a plasmonic nanostructure}
\label{sec:emission}

In the following, we consider the emission spectrum for a \textit{single} QE situated near a plasmonic structure. If the QE emission frequency $\omega_{e}$ is close to a plasmon mode frequency $\omega_{m}$, we can restrict ourselves to a single resonant plasmon mode. To highlight the role of dispersion-induced non-Markovian effects, we consider the case when only a \textit{single} excitation is present in the hybrid system and disregard any nonlinear effects. Furthermore, surface plasmons normally interact with excitons in quantum dots or dye molecules which can be approximately described, if one ignores  exciton's internal structure, by simple bosonic operators  $e$ and $e^{\dagger}$ obeying $[e,e^{\dagger}]=1$. As we show below, even in this simple case, non-Markovian effects due to optical dispersion of the metal dielectric function can dramatically affects the emission spectrum as the system transitions to strong coupling regime.

\subsection{The Hamiltonian and Heisenberg equations}

Within quantum approach developed in the previous section, in the absence of external field, the system is described by the   Hamiltonian
\begin{align}
\label{H-prm-qe}
\hat{H}=   \!\int_{0}^{\infty}\!   d\omega \,\hbar\omega \,\hat{b}_{m}^{\dagger}(\omega)\hat{b}_{m}(\omega)
+\hbar\omega_{e}\hat{e}^{\dagger}\hat{e}
~~~~~~~~~~~~~~~~
\nonumber
\\
+  \int_{0}^{\infty}\!  d\omega\left [\hbar q_{em}(\omega)\hat{e}^{\dagger}\,\hat{b}_{m}(\omega) +\text{H.c.}\right ],
 \end{align} 
where  $q_{em}(\omega)=g_{em}\lambda_{m}(\omega)$ is the QE-PRM coupling. Here,  $g_{em}=-\bm{\mu}_{e}\!\cdot\!\tilde{\bm{E}}_{m}(\bm{r}_{e})/\hbar$ is the QE-plasmon coupling  that, in terms of plasmon mode volume, is given by  Eq.~(\ref{coupling-mode-volume-main}), while $\lambda_{m}(\omega)$  contains the plasmon pole [see Eq.~(\ref{lambda-main})].  Hereafter, for a single QE, the subscript $i$ is replaced with $e$ in all expressions. Using the Hamiltonian (\ref{H-prm-qe}), the Heisenberg equation for the PRM operator is obtained as
\begin{align}
\label{HE-b}
i\dot{\hat{b}}_{m}(t,\omega)  =\omega \hat{b}_{m}(t,\omega) +  q^{*}_{em}(\omega)\hat{e}(t),
\end{align}
while for QE operator, the Heisenberg equation is
\begin{align}
\label{HE-qe}
i\dot{\hat{e}}(t)  =(\omega_{e} -i\gamma_{e}/2)\hat{e}(t) 
+\!\int_{0}^{\infty}\!  d\omega \,  q_{em}(\omega) \hat{b}_{m}(t,\omega),
\end{align}
where we added the decay rate  $\gamma_{e}$ of an isolated QE.

We assume that at $t=0$, only the QE is excited. Introducing the Laplace transforms of QE and PRM operators, respectively, as $\hat{e}(\omega)=-i\int_{0}^{\infty}dt e^{i\omega t}\hat{e}(t)$ and $\hat{b}_{m}(\omega,\omega')=-i\int_{0}^{\infty}dt e^{i\omega t}\hat{b}_{m}(t,\omega')$, where an infinitesimal positive imaginary part of $\omega$ is implied, we integrate Eqs.~(\ref{HE-b}) and (\ref{HE-qe}) to obtain
\begin{equation}
\label{b-laplace}
\hat{b}_{m}(\omega,\omega')=\frac{q_{em}^{*}(\omega')\hat{e}(\omega)}{\omega-\omega'} 
\end{equation}
and, correspondingly,
\begin{align}
\label{qe-laplace}
\hat{e}(\omega)=\frac{1}{\Omega_{e}(\omega)}\left [\hat{e}_{0}+\!\int_{0}^{\infty}\! \! d\omega'    q_{em}(\omega') \hat{b}_{m}(\omega,\omega')\right ],
\end{align}
where $\hat{e}_{0}\equiv \hat{e}(t=0)$. For brevity, we use the notations
\begin{equation}
\Omega_{e}(\omega)=\omega-\omega_{e}+\frac{i}{2}\gamma_{e},
~~
\Omega_{m}(\omega)=\omega-\omega_{m}+\frac{i}{2}\gamma_{m}(\omega).
\end{equation}
Eliminating $\hat{b}_{m}$ from Eq.~(\ref{qe-laplace}) and employing the relation
\begin{equation}
\label{kk}
\int_{0}^{\infty}\!\!  d\omega' \, \frac{|\lambda_{m}(\omega')|^{2}}{\omega-\omega'+i0} 
%=\frac{1}{\pi} \int_{0}^{\infty}\!\! d\omega' \, \frac{\text{Im}\, \Omega_{m}^{-1}(\omega')}{\omega-\omega'}
=\frac{1}{\Omega_{m}(\omega)},
\end{equation}
where we used that $|\lambda_{m}(\omega)|^{2}=\pi^{-1}\text{Im}\, \Omega_{m}^{-1}(\omega)$, we obtain a closed equation for $\hat{e}(\omega)$
\begin{align}
\label{qe-laplace2}
\hat{e}(\omega)=\frac{1}{\Omega_{e}(\omega)}
\left [ \hat{e}_{0}+ \frac{ g_{em}^{2}}{\Omega_{m}(\omega)}\, \hat{e}(\omega) \right ].
\end{align}
Note that, strictly speaking, the real part of Eq.~(\ref{kk}) (i.e., the Kramers-Kronig relation) should include the negative frequency contribution  as well; however, within the RWA, this contribution is neglected. Solving Eq.~(\ref{qe-laplace2}), we obtain the QE operator as
\begin{align}
\label{qe-laplace3}
\hat{e}(\omega)
%=\frac{1}{\Omega_{0}(\omega)}\left[1+
%\frac{g^{2}}{\Omega_{e}(\omega)\Omega_{m}(\omega) -g_{em}^{2}}\right]\hat{c}_{0}
%\nonumber
%\\
=\frac{\Omega_{m}(\omega)\,\hat{e}_{0}}{\Omega_{e}(\omega)\Omega_{m}(\omega) -g_{em}^{2}},
\end{align}
which, in turn, defines the  PRM operator via Eq.~(\ref{b-laplace}).

\subsection{System dipole moment and emission spectrum}

The  hybrid plasmonic system interacts with the EM field via the  dipole moment operator $\hat{\bm{p}}_{s}=\hat{\bm{p}}_{e}+\hat{\bm{p}}_{m}$  that includes both QE and PRM contributions (in the RWA). From Eq.~(\ref{qe-laplace3}), the QE dipole moment  readily follows as $\hat{\bm{p}}_{e}(\omega)=\bm{\mu}_{e} \hat{e}(\omega)=\bm{p}_{e}(\omega)\hat{e}_{0}$, where
\begin{align}
\label{qe-dipole}
\bm{p}_{e}(\omega)
=\frac{\bm{\mu}_{e}\Omega_{m}(\omega)}{\Omega_{e}(\omega)\Omega_{m}(\omega) -g_{em}^{2}}.
\end{align}
Consider now the RPM dipole operator [see Eq.~(\ref{H-b-em-main})]
\begin{equation}
\label{prm-dipole}
\hat{\bm{p}}_{m}(\omega)=\int_{0}^{\infty} \!\! d\omega'  \bm{d}_{m} (\omega,\omega')\hat{b}_{m}(\omega,\omega'),  
\end{equation}
where $\bm{d}_{m}(\omega,\omega')=\bm{\mu}_{m}(\omega)\lambda_{m}(\omega')$. Using Eqs.~(\ref{b-laplace}), (\ref{kk}) and (\ref{qe-laplace3}),  we  obtain $\hat{\bm{p}}_{m}(\omega)= \bm{p}_{m}(\omega)\hat{e}_{0}$, where
\begin{equation}
\label{prm-dipole2}
\bm{p}_{m}(\omega)
=\frac{g_{em}\bm{\mu}_{m}(\omega)}{\Omega_{e}(\omega)\Omega_{m}(\omega) -g_{em}^{2}}.
\end{equation}
Combining Eqs.~(\ref{qe-dipole}) and (\ref{prm-dipole2}), we obtain the system dipole moment operator as $\hat{\bm{p}}_{s}(\omega)= \bm{p}_{s}(\omega)\hat{e}_{0}$, where
\begin{align}
\label{system-dipole}
\bm{p}_{s}(\omega)
%=\bm{p}_{e}(\omega)+\bm{p}_{m}(\omega)
=\frac{\bm{\mu}_{e}\Omega_{m}(\omega)+ g_{em}\bm{\mu}_{m}(\omega)}{\Omega_{e}(\omega)\Omega_{m}(\omega) -g_{em}^{2}}.
\end{align}

Turning to the emission spectrum, we  note that if the characteristic system size is smaller than the radiation wavelength then system's interaction with light  can be treated similar to that of a localized dipole $\hat{\bm{p}}_{s}$ situated at some point  $\bm{r}_{s}$, which, for convenience, we set at the origin. The far field  generated by such a  dipole is $\hat{\bm{E}}_{s}(\omega,\bm{r})=\bm{D}_{0}(\omega;\bm{r})\hat{\bm{p}}_{s}(\omega)$, where $\bm{D}_{0}(\omega;\bm{r})$ is the free-space EM Green function.  The spectral intensity of emitted light detected at a remote point $\bm{r}$ is defined as \cite{andreani-prb12} $S(\omega,\bm{r})=(c/4\pi^{2})\langle \hat{\bm{E}}_{s}(\omega,\bm{r})\!\cdot\!\hat{\bm{E}}_{s}^{\dagger}(\omega,\bm{r})\rangle$, which, using the above relation, takes the form
\begin{equation}
S(\omega,\bm{r})
=\frac{c}{4\pi^{2}}\left |\bm{D}_{0}(\omega;\bm{r})\right |^{2}\langle \hat{\bm{p}}_{s}(\omega)\!\cdot\!\hat{\bm{p}}_{s}^{\dagger}(\omega)\rangle.
\end{equation}
The full radiated energy in  unit frequency interval is obtained by integrating  $S(\omega,\bm{r})$ over a remote spherical surface: $S(\omega)=r^{2}\!\int\! d\Omega S(\omega;\bm{r})$ at $r\rightarrow \infty$ ($\Omega$  is solid angle). Using far-field asymptotics of the free-space EM Green function $\bm{D}_{0}(\omega;\bm{r})=(\omega^{2}/c^{2}r)\left (\bm{I}-\hat{\bm{r}}\hat{\bm{r}}\right )e^{ikr}$ \cite{novotny-book} and the relation $\int\! d\Omega \,\hat{\bm{r}}\hat{\bm{r}}=(4\pi/3) \bm{I}$, where $\hat{\bm{r}}$ is the unit vector, it is straightforward to obtain
\begin{equation}
S(\omega)=\frac{2\omega^{4}}{3\pi c^{3}}\,\langle \hat{\bm{p}}_{s}(\omega)\!\cdot\!\hat{\bm{p}}_{s}^{\dagger}(\omega)\rangle
=\frac{2\omega^{4}}{3\pi c^{3}}\left |\bm{p}_{s}(\omega)\right |^{2}.
\end{equation}
Finally, using Eq.~(\ref{system-dipole}), we  obtain the emission spectrum:
\begin{align}
\label{spectrum}
S(\omega)
%=\frac{2\omega^{4}}{3\pi c^{3}}\left | \bm{p}_{e}(\omega)+\bm{p}_{m}(\omega)\right |^{2}
=\frac{2\omega^{4}}{3\pi c^{3}}\left |\frac{\bm{\mu}_{e}\Omega_{m}(\omega)+ g_{em}\bm{\mu}_{m}(\omega)}{\Omega_{e}(\omega)\Omega_{m}(\omega) -g_{em}^{2}} \right |^{2}.
\end{align}
The above expression is valid for a plasmonic structure of arbitrary shape and composition provided that its characteristic size is below the diffraction limit. Importantly, the emission spectrum (\ref{spectrum}) accurately accounts for the metal dielectric function's dispersion incorporated in the plasmon's induced dipole moment $\bm{\mu}_{m}(\omega)$ and the decay rate $\gamma_{m}(\omega)$. Below we demonstrate that dispersion-induced non-Markovian effects strongly affect the emission spectrum in the strong coupling regime.

\section{Weak coupling regime: Dressed emitter vs. plasmonic antenna}
\label{sec:weak}

We start with an analysis of the weak coupling regime. In this regime, there are no distinct polaritonic bands, and so the emission spectrum (\ref{spectrum}), which includes the QE and plasmon contributions,  represents a single peak. Consider first the QE part, $S_{e}(\omega)=(2\omega^{4}/3\pi c^{3})\left | \bm{p}_{e}(\omega)\right |^{2}$, where  $\bm{p}_{e}(\omega)$  can be  presented as
\begin{align}
\label{qe-dipole3}
\bm{p}_{e}(\omega)
=\frac{\bm{\mu}_{e}}{\Omega_{e}(\omega)-g_{em}^{2}/\Omega_{m}(\omega)}
=\frac{\bm{\mu}_{e}}{\omega-\omega'_{e}(\omega)+\frac{i}{2} \Gamma_{e}(\omega)}.
\end{align}
Here $\Gamma_{e}(\omega)=\gamma_{e}+\gamma_{e\rightarrow m}(\omega)$ is the QE  full decay rate that  includes the QE-plasmon ET rate $\gamma_{e\rightarrow m}(\omega)$, given by Eq.~(\ref{rate-et-mode-main}), while $\omega'_{e}(\omega)=\omega_{e}+\delta\omega_{e}(\omega)$ is the interacting QE's  frequency that now includes the frequency shift due to  near-field coupling between the QE and  plasmon,
\begin{equation}
\label{shift-qe-mode}
\delta\omega_{e}(\omega)=\frac{g_{em}^{2} (\omega-\omega_{m})}{(\omega_{m}-\omega)^{2} +\frac{1}{4}\gamma_{m}^{2}(\omega)}.
\end{equation}
Then  the QE  contribution to the emission spectrum can be written in the form
\begin{equation}
\label{spectrum-qe}
S_{e}(\omega)=\frac{\hbar\omega}{2} \frac{\eta_{e}(\omega) \Gamma_{e}(\omega)}{\left | \omega-\omega'_{e}(\omega)+\frac{i}{2} \Gamma_{e}(\omega)\right |^{2}},
\end{equation}
where $\eta_{e}(\omega)=\gamma_{e}^{rad}(\omega)/\Gamma_{e}(\omega)$ is  QE's radiation efficiency; here, $\gamma_{e}^{rad}(\omega)=4\mu_{e}^{2}\omega^{3}/3\hbar c^{3}$ is its free-space radiative decay rate. Note that, in dye molecules or quantum dots, optical transitions are accompanied by phonon or vibron excitation and, therefore, the  emission spectrum (\ref{spectrum-qe}) should be integrated with the corresponding distribution function, so that, in the weak coupling regime, the effective rate  $\gamma_{e}$ is typically much greater than the frequency shift $\delta\omega_{e}$ \cite{novotny-book}.  Finally, the energy $W_{e}$ radiated by the dressed QE   is obtained by  integrating $S_{e}(\omega)$ over the entire frequency range,
\begin{equation}
\label{energy-emitter}
W_{e}=\int\!d\omega S_{e}(\omega)=\hbar\omega_{e} \eta_{e}(\omega_{e}),
\end{equation}
implying that, near the plasmonic structure, $W_{e}$ is reduced due to ET from the QE to  plasmon. 

Let us turn to plasmon's contribution to the emission spectrum $S_{m}(\omega)=(2\omega^{4}/3\pi c^{3})\left | \bm{p}_{m}(\omega)\right |^{2}$,  where $\bm{p}_{m}(\omega)$ is given by Eq.~(\ref{prm-dipole2}). In a similar way, using Eq.~(\ref{rate-et-mode-main}), we get
\begin{equation}
\label{spectrum-mode}
S_{m}(\omega)=\frac{\hbar\omega}{2} \frac{\eta_{m}(\omega) \eta_{e\rightarrow m}(\omega) \Gamma_{e}(\omega)}{\left | \omega-\omega'_{e}(\omega)+\frac{i}{2} \Gamma_{e}(\omega)\right |^{2}},
\end{equation}
where $\eta_{m}(\omega)=\gamma_{m}^{rad}(\omega)/\gamma_{m}(\omega)$ is the plasmon radiation efficiency and $\eta_{e\rightarrow m}(\omega)=\gamma_{e\rightarrow m}(\omega)/\Gamma_{e}(\omega)$ is the QE-plasmon ET efficiency. This contribution describes the \textit{plasmonic antenna} effect as the energy is transferred from the QE to plasmon with efficiency $\eta_{e\rightarrow m}(\omega)$ and then radiated away  with efficiency $\eta_{m}(\omega)$. Accordingly, the full energy radiated by the plasmonic antenna is determined by their product, 
\begin{equation}
\label{energy-antenna}
W_{m}=\int\!d\omega S_{e}(\omega)=\hbar\omega_{e} \eta_{m}(\omega_{e}) \eta_{e\rightarrow m}(\omega_{e}).
\end{equation}
To estimate the relative importance of dressed QE's and plasmomic antenna's contributions, we use Eqs.~(\ref{energy-emitter}) and (\ref{energy-antenna}) to present the ratio of their radiated energies as 
\begin{equation}
\frac{W_{m}}{W_{e}}=F_{p} \eta_{m},
\end{equation}
where $F_{p}=\gamma_{e\rightarrow m}/\gamma_{e}^{rad}$ is the Purcell factor that characterizes  ET from QE to plasmonic antenna. For large Purcell factors  indicating a highly efficient QE-plasmon ET, the light mainly emanates from the  antenna due to its much greater dipole moment. 

Note that in the Markov approximation and with phenomenological coupling parameters, a similar expression for the antenna spectrum (\ref{spectrum-mode}) can  obtained within a classical coupled-oscillator (CO) model with one of the oscillators (QE) considered  dark due to its much smaller dipole moment \cite{pelton-oe10}.  In addition to the QE and plasmon contributions discussed above, the emission spectrum (\ref{spectrum}) also includes the interference  between the plasmon and plasmon-induced dipole moments. Although the QE dipole moment is typically much smaller than the plasmon one, for small nanostructures characterized by small antenna size, such an interference does affect the spectral weight of polaritonic bands in the scattering spectra \cite{shahbazyan-nanophot21}. In the following section, we demonstrate that  non-Markovian effects  caused by  the dielectric function dispersion can dramatically  affect the shape of emission spectrum in the strong coupling regime.

\section{Strong coupling regime and non-Markovian effects}
\label{sec:strong}
%\subsection{Strong coupling regime: Emission spectrum shape and non-Markovian effects}

%%%%%%%%%%%%%%%%%%%%%%%%%%%%%%%%%%%%%%%%%%%%%%
%
\begin{figure}[tb]
%\centering
\begin{center}
\includegraphics[width=0.85\columnwidth]{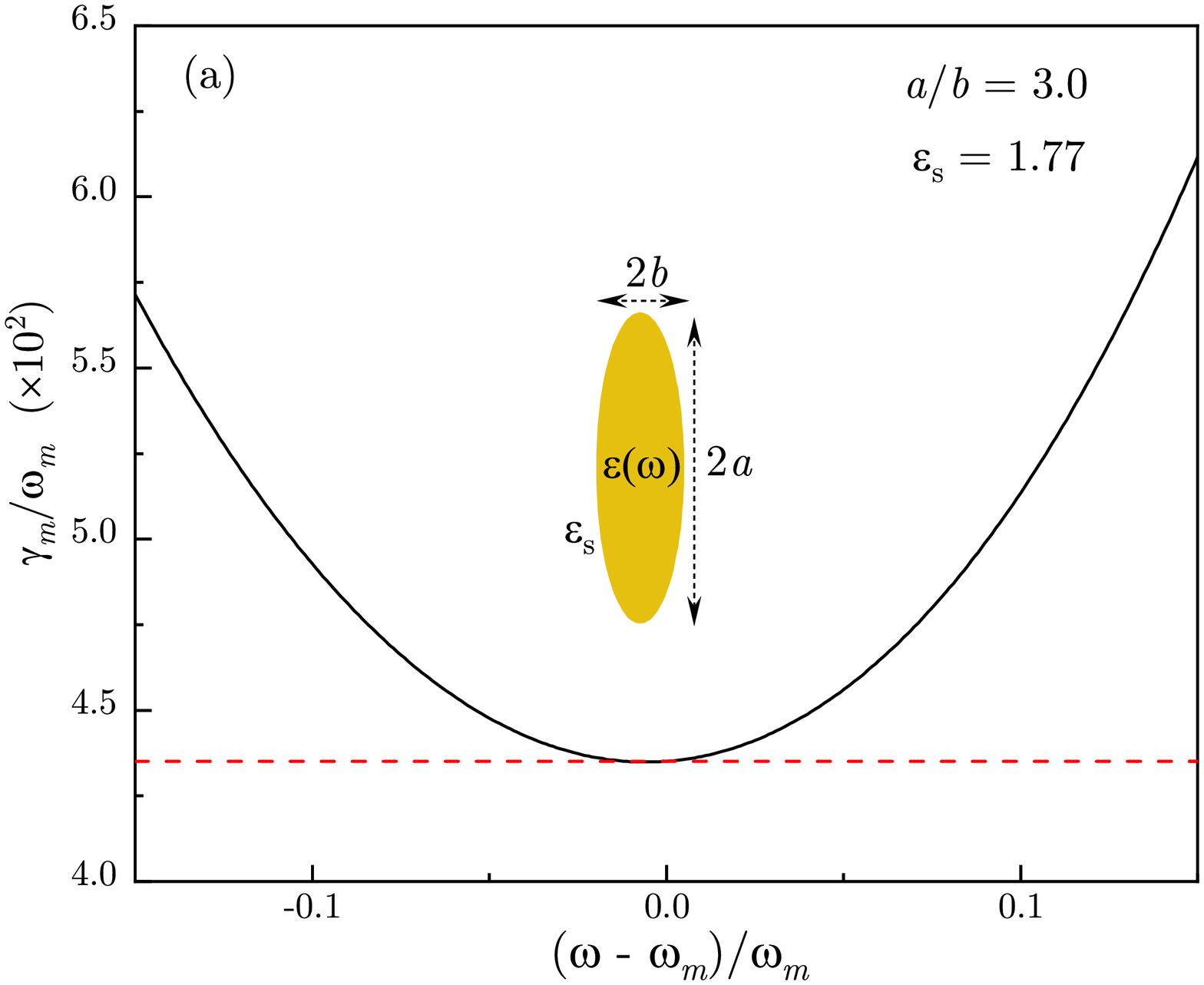}

\vspace{2mm}

\includegraphics[width=0.85\columnwidth]{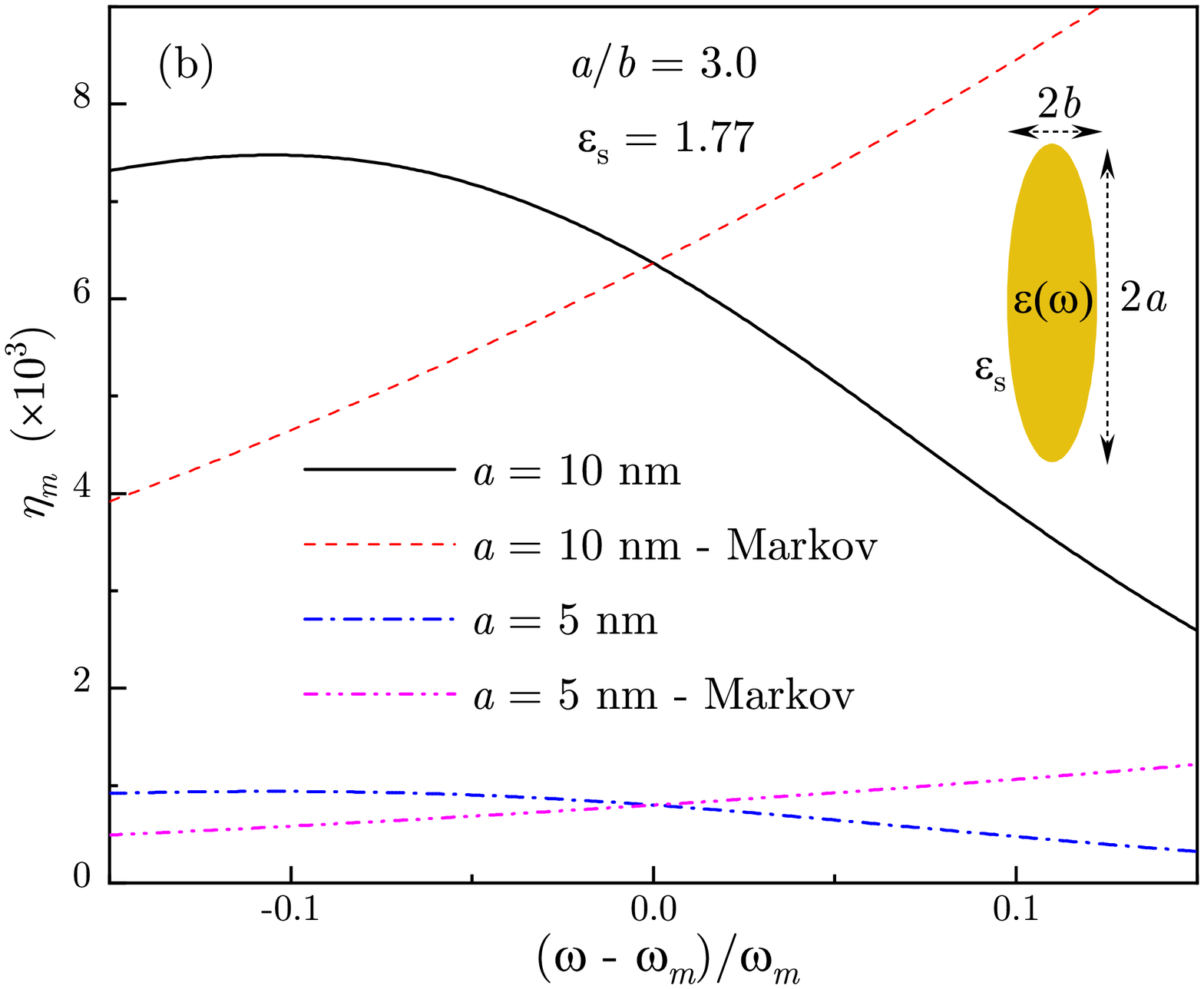}
\caption{\label{fig1} (a) Frequency-dependent plasmon's decay rate is compared with its Markov approximation value (dashed line). (b) Frequency-dependent plasmon's radiation efficiency for nanorod sizes $2a=20$ nm and  $2a=10$ nm is compared with their Markov approximation counterparts. Inset shows schematics of a Au nanorod in water.
 }
\end{center}
\vspace{-4mm}
\end{figure}
%
%%%%%%%%%%%%%%%%%%%%%%

\begin{figure}[tb]
%\centering
\begin{center}
\includegraphics[width=0.85\columnwidth]{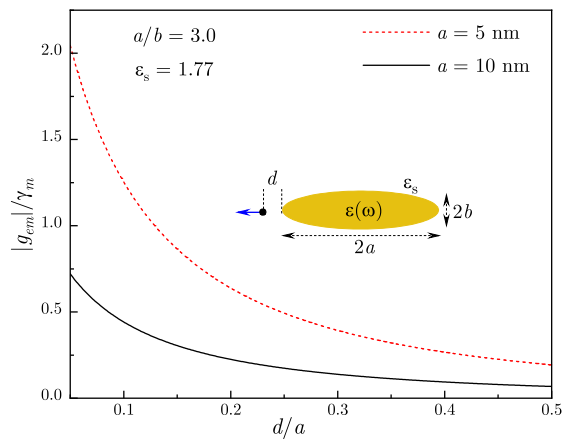}
\caption{\label{fig2} Normalized QE-plasmon coupling  is plotted against  QE distance $d$ to the Au nanorod tip for nanorod sizes $2a=20$ nm and  $2a=10$ nm. Inset: Schematics of a QE situated near a Au nanorod in water.
 }
\end{center}
\vspace{-3mm}
\end{figure}

To describe the transition to strong coupling regime, we perform numerical calculations for a QE situated at a distance $d$ from the tip of an Au nanorod in water with excitation frequency in resonance with the surface plasmon frequency, i.e., $\omega_{e}=\omega_{m}$. The nanorod was modeled by a prolate spheroid  with semimajor and semiminor axes $a$ and $b$, respectively (see insets in Figs. \ref{fig1} and \ref{fig2}). Calculations were carried out for two nanorods of overall length $2a=20$ nm and $2a=10$ nm, while the aspect ratio was fixed at $a/b=3.0$  to ensure the same longitudinal plasmon  frequency $\omega_{m}$ for both nanorods. The orientation of QE's dipole moment $\bm{\mu}_{e}$ was chosen along the nanorod symmetry axis,  and  the Au experimental dielectric function $\varepsilon(\omega)=\varepsilon'(\omega)+i\varepsilon''(\omega)$ was used throughout \cite{johnson-christy} (the dielectric constant of water is $\varepsilon_{s}=1.77$). For these parameters, the plasmon resonance  wavelength is  $\lambda_{m}\approx 675$ nm, at which $\varepsilon''(\omega)$ reaches its smallest value; this, in turn, translates to maximal value of the local density of states (LDOS) \cite{shahbazyan-prl16} that determines the QE-plasmon coupling strength \cite{shahbazyan-nl19}. We employed the standard spherical harmonics formalism for calculation of local fields near the tip  of a prolate spheroid which, together with $\varepsilon(\omega)$,  determine the plasmon parameters  $\mu_{m}(\omega)$, $\gamma_{m}(\omega)$, $\eta_{m}(\omega)$ and the QE-plasmon coupling $g_{em}$. The QE spectral linewidth $\gamma_{e}$ was chosen as $\gamma_{e}/\gamma_{m}=0.1$ relative to the plasmon decay rate $\gamma_{m}$, and the QE radiative decay time was chosen $\tau_{e}^{rad}= 10$ ns, which are typical values for excitons in  quantum dots \cite{pelton-nc18}. Note that the  QE radiative decay rate $\gamma_{e}^{rad}$ is much smaller that its spectral linewidth, which contains phonon or vibron contributions: for our system we have $\gamma_{e}^{rad}/\gamma_{e}\approx 10^{-5}$.

%%%%%%%%%%%%%%%%%%%%%%%%%%%%%%%%%%%%%%%%%%%%%%
%
%
%%%%%%%%%%%%%%%%%%%%%%

%%%%%%%%%%%%%%%%%%%%%%%%%%%%%%%%%%%%%%%%%%%%%%
%
\begin{figure}[tb]
%\centering
\begin{center}
\includegraphics[width=0.85\columnwidth]{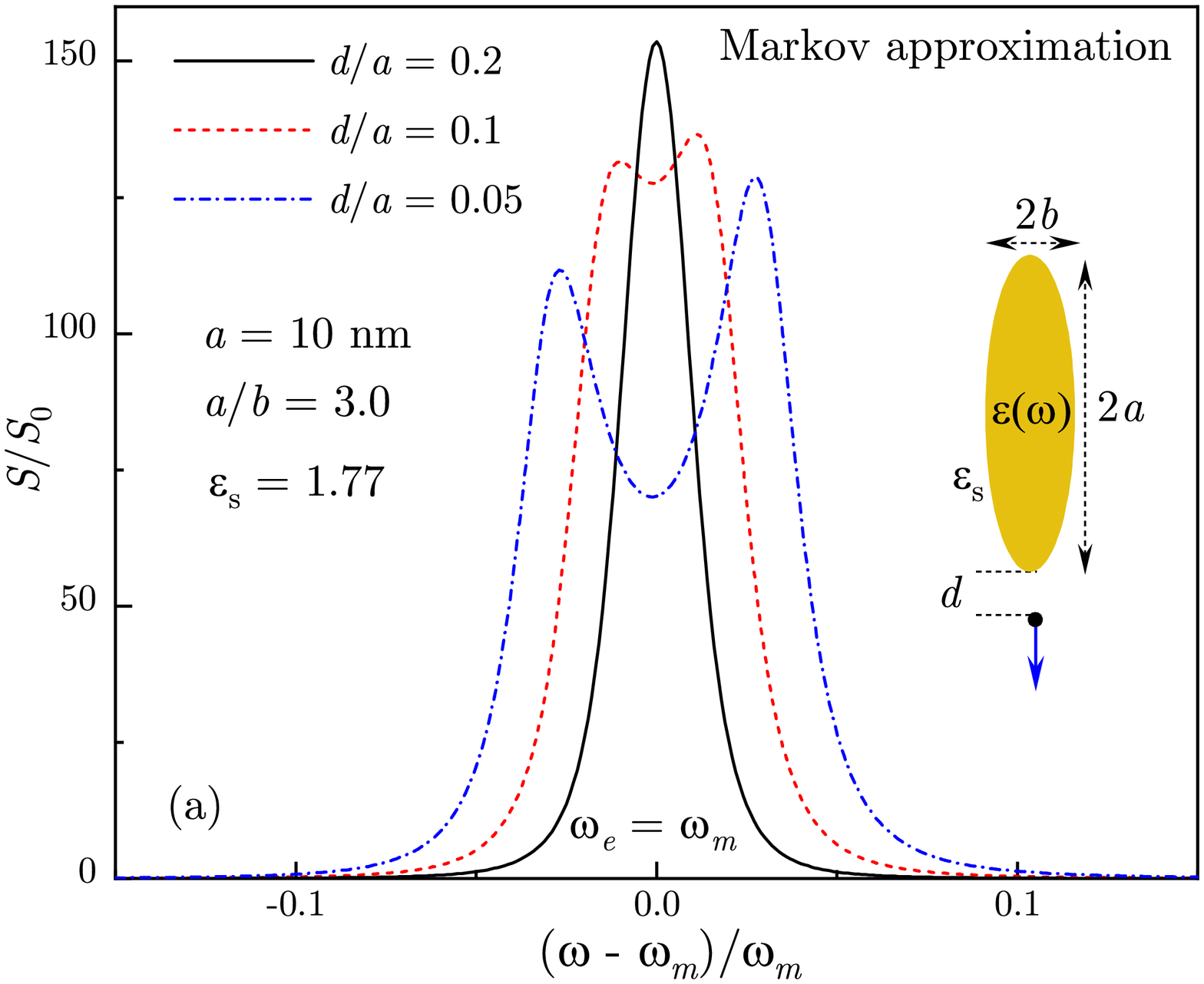}

\vspace{2mm}

\includegraphics[width=0.85\columnwidth]{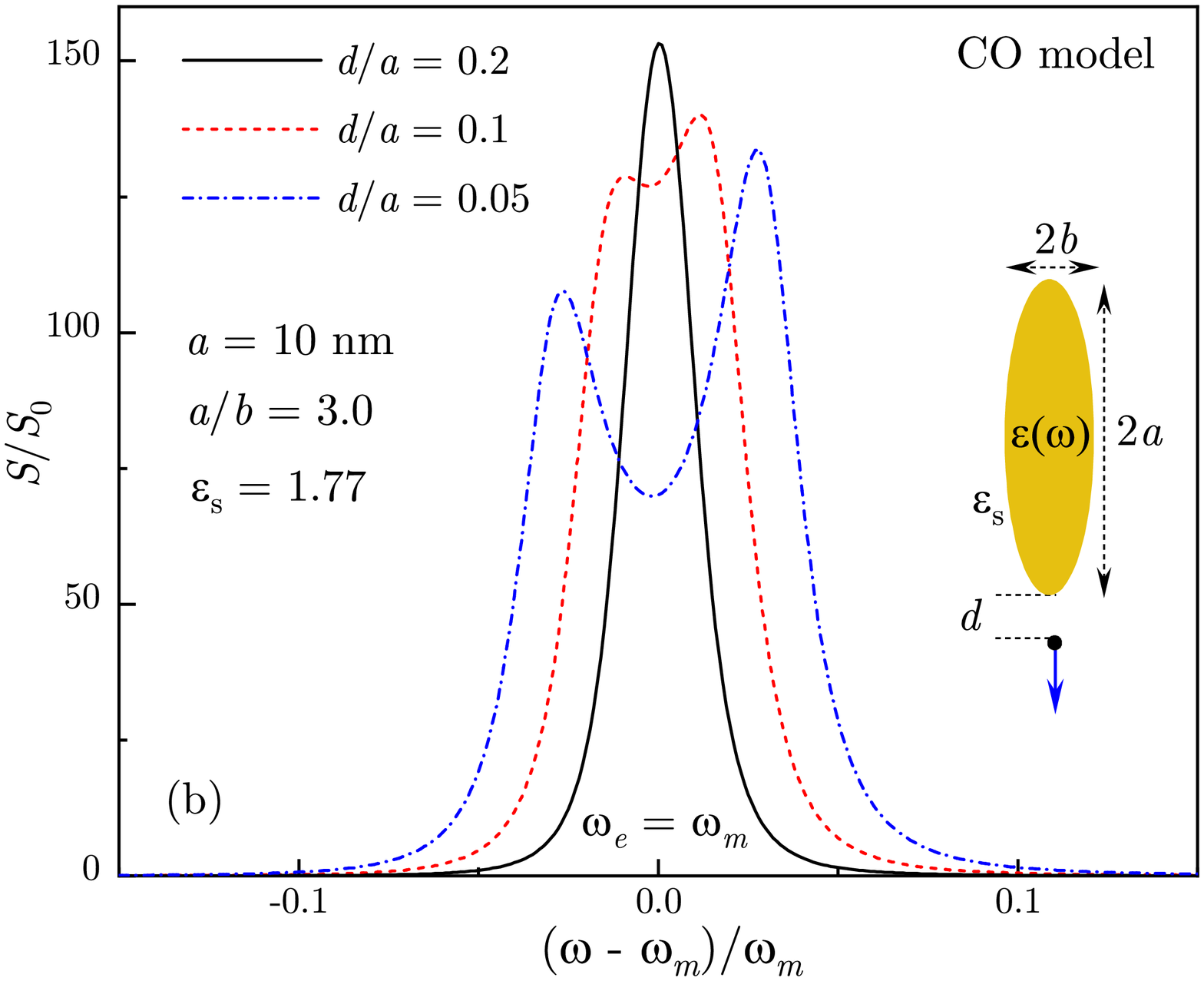}

\caption{\label{fig3} Normalized  emission spectra for nanorod size $2a=20$ nm obtained in the Markov approximation are shown at several values of $d$ for (a)   plasmonic system with both plasmon and QE dipoles coupled to the radiation field and (b) CO model with only plasmon dipole coupled to the radiation field.  
 }
\end{center}
\vspace{-4mm}
\end{figure}
%
%%%%%%%%%%%%%%%%%%%%%%

%%%%%%%%%%%%%%%%%%%%%%%%%%%%%%%%%%%%%%%%%%%%%%
%
\begin{figure}[t]
%\centering
\begin{center}
\includegraphics[width=0.85\columnwidth]{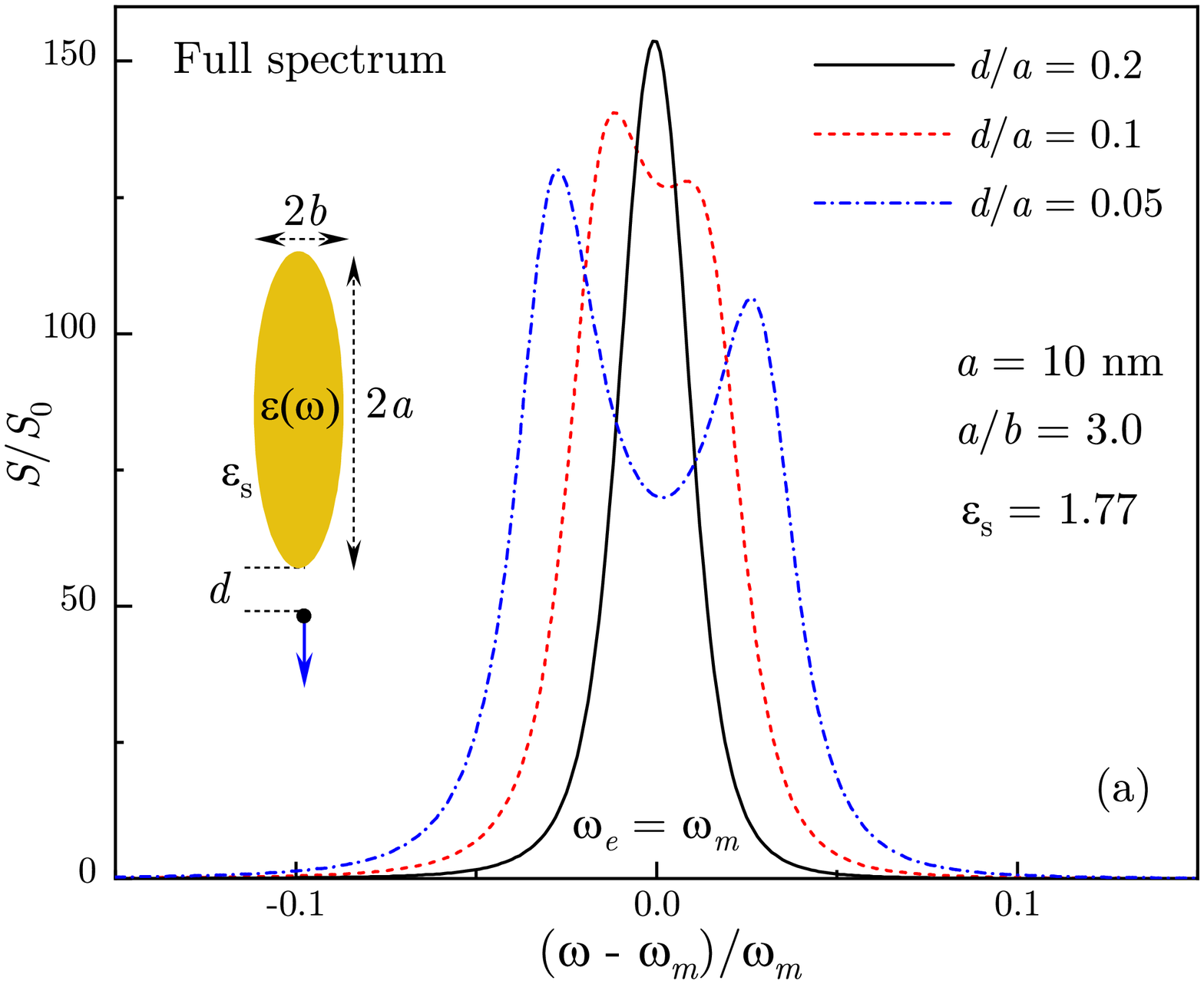}

\vspace{2mm}

\includegraphics[width=0.85\columnwidth]{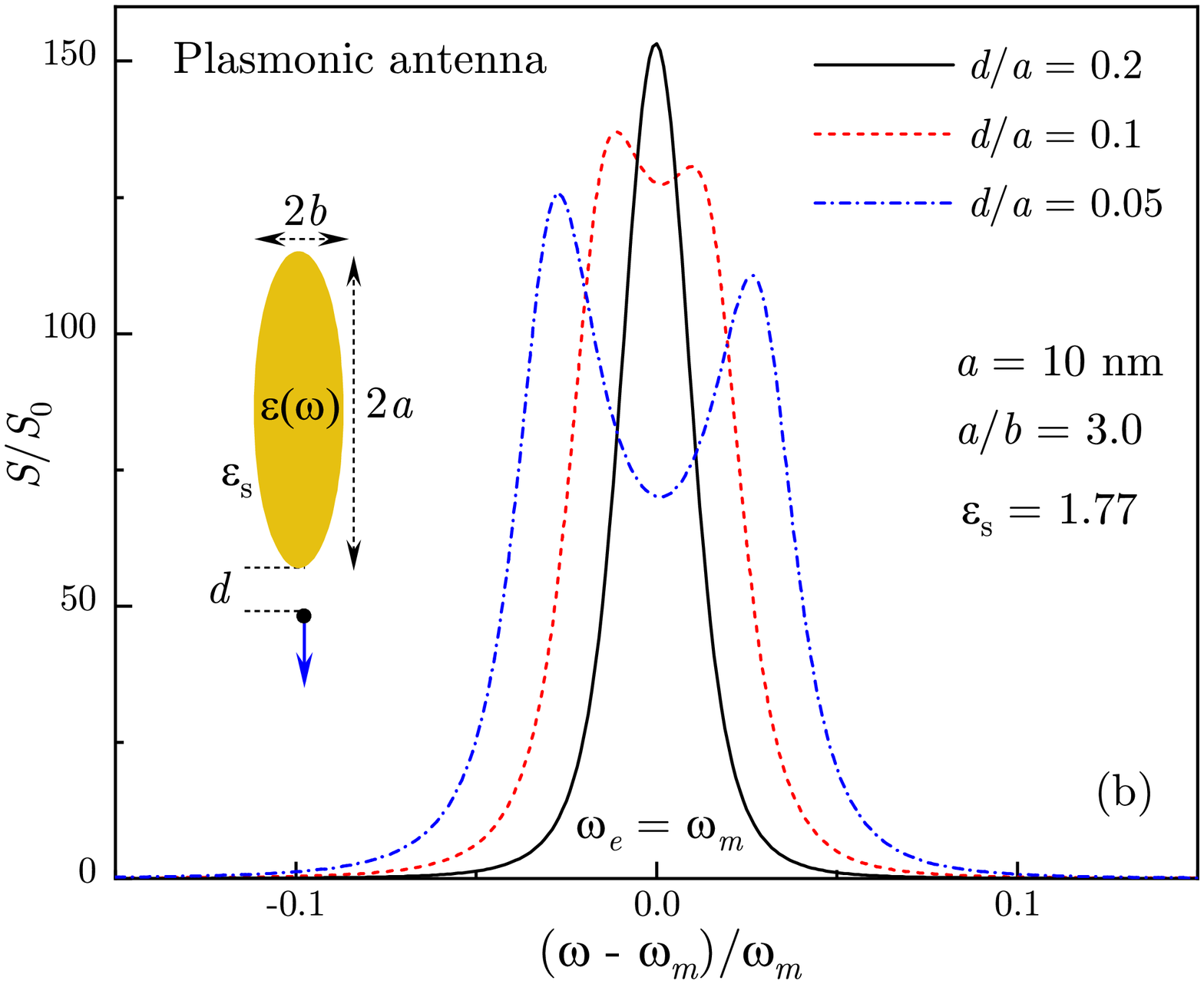}

\caption{\label{fig4} Normalized  emission spectra  for nanorod size $2a=10$ nm are shown at several values of $d$ for (a)  plasmonic system with both  plasmon and QE coupled to the radiation field and (b)  plasmonic antenna contribution only. Inset shows schematics of a QE  near the Au nanorod tip.
 }
\end{center}
\vspace{-4mm}
\end{figure}
%
%%%%%%%%%%%%%%%%%%%%%%

%%%%%%%%%%%%%%%%%%%%%%%%%%%%%%%%%%%%%%%%%%%%%%
%
\begin{figure}[t]
%\centering
\begin{center}
\includegraphics[width=0.85\columnwidth]{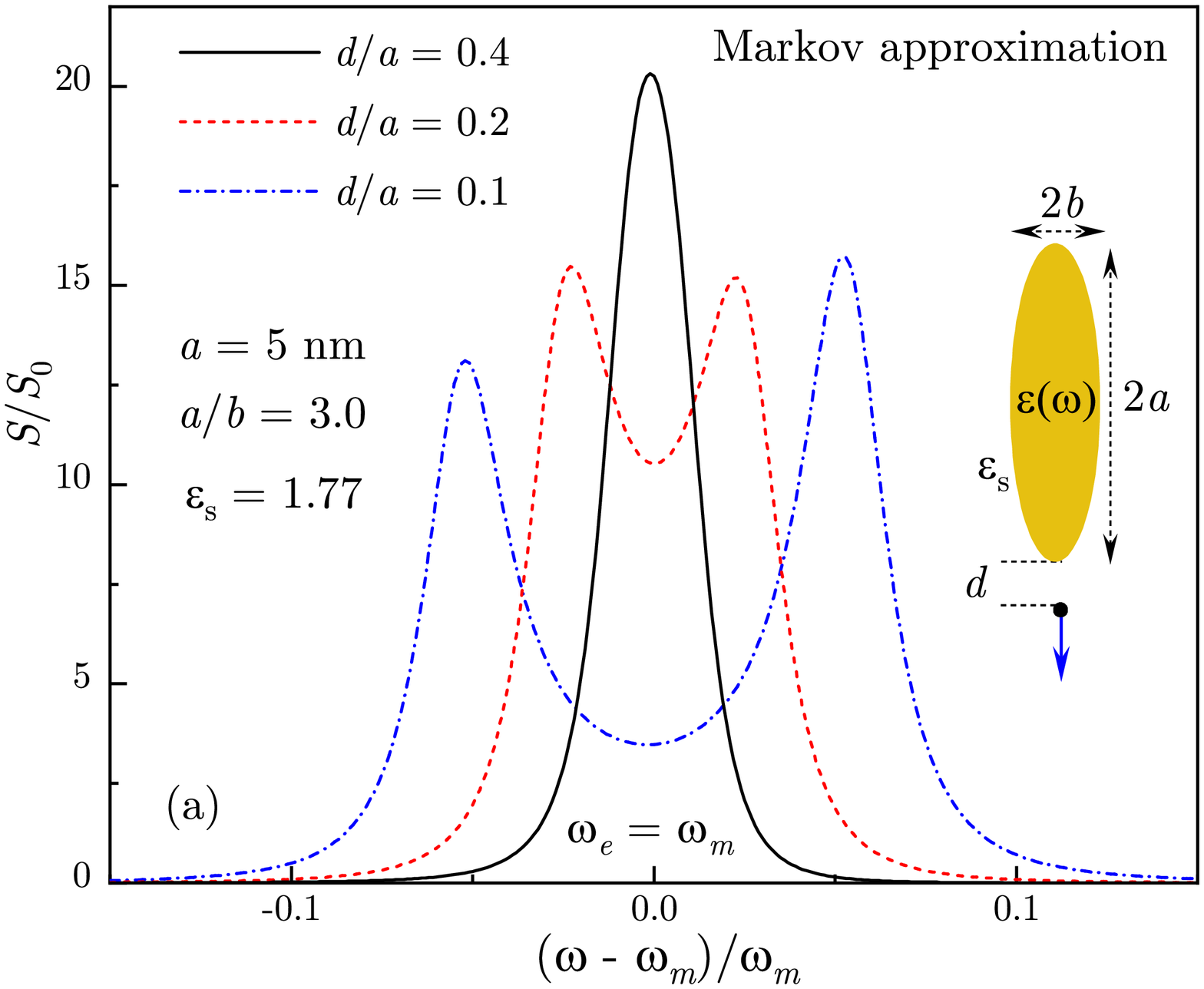}

\vspace{2mm}

\includegraphics[width=0.85\columnwidth]{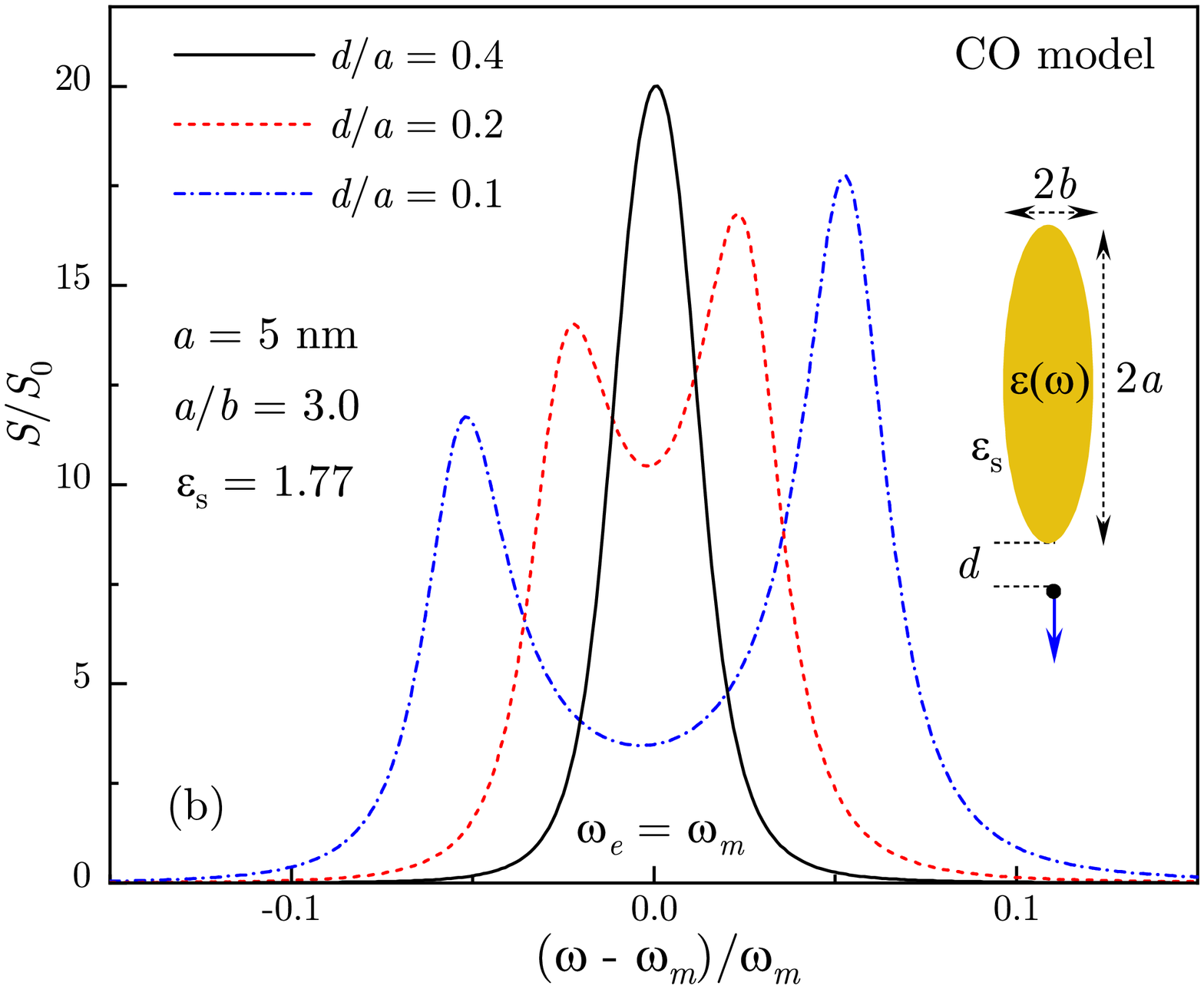}

\caption{\label{fig5} Normalized  emission spectra for nanorod size $2a=10$ nm obtained in the Markov approximation are shown at several values of $d$ for (a)  plasmonic system with both plasmon and QE coupled to the radiation field and (b) CO model with only plasmon dipole coupled to the radiation field.  
 }
\end{center}
\vspace{-4mm}
\end{figure}
%
%%%%%%%%%%%%%%%%%%%%%%

%%%%%%%%%%%%%%%%%%%%%%%%%%%%%%%%%%%%%%%%%%%%%%
%
\begin{figure}[t]
%\centering
\begin{center}
\includegraphics[width=0.85\columnwidth]{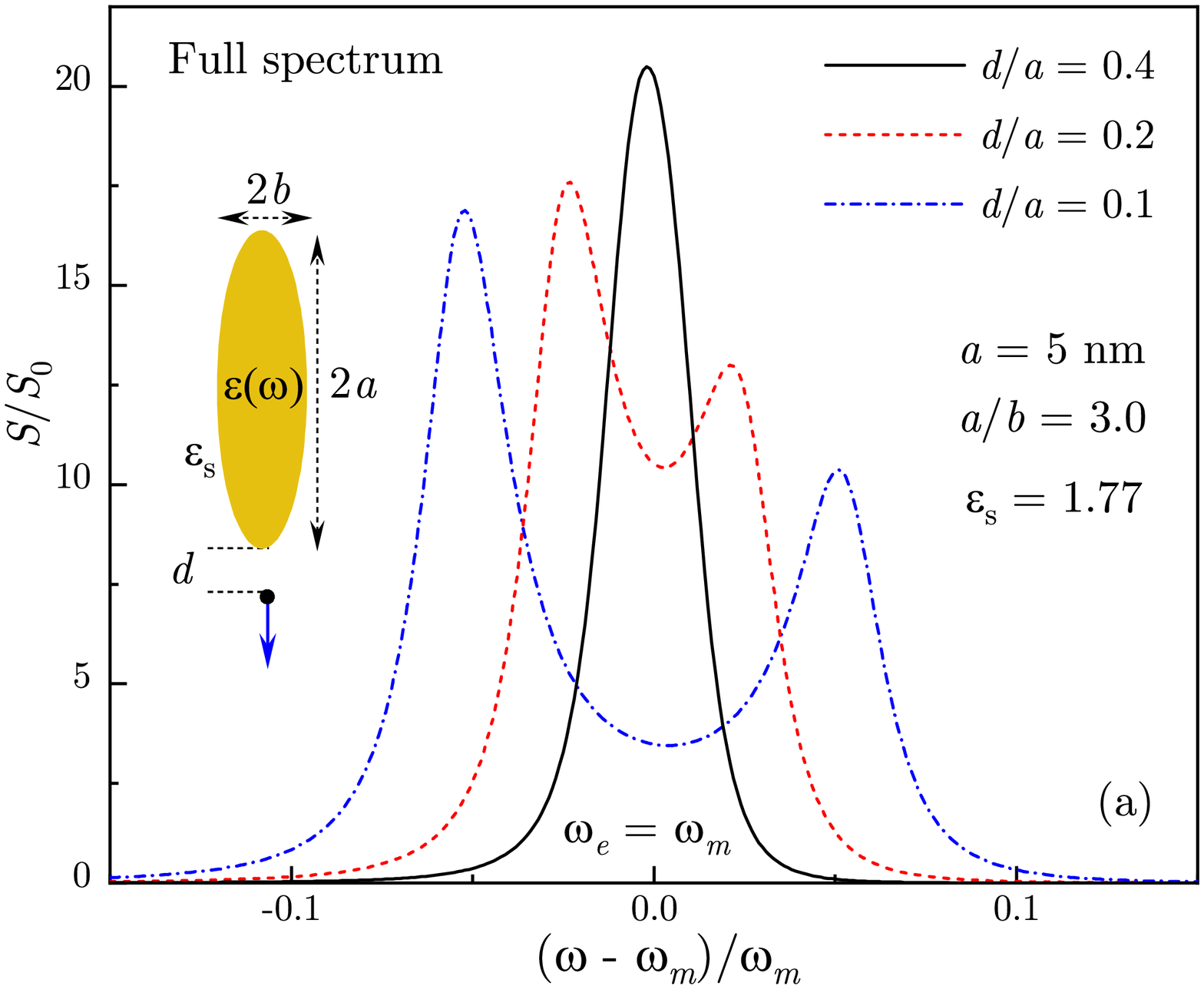}

\vspace{2mm}

\includegraphics[width=0.85\columnwidth]{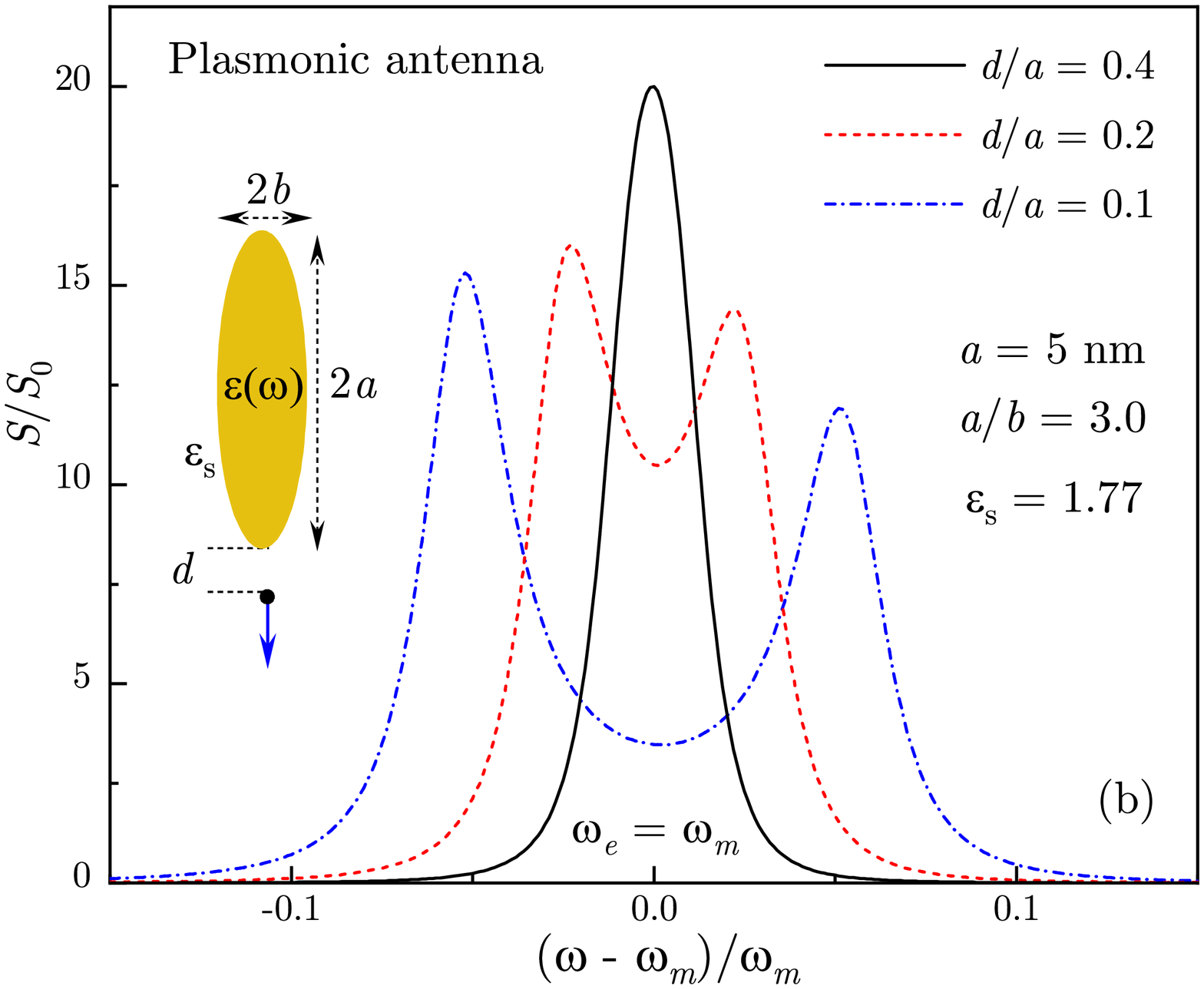}

\caption{\label{fig6} Normalized  emission spectra  for nanorod size $2a=10$ nm are shown at several values of $d$ for (a)  plasmonic system with both  plasmon and QE coupled to the radiation field and (b)  plasmonic antenna contribution only. Inset shows schematics of a QE  near the Au nanorod tip.
 }
\end{center}
\vspace{-4mm}
\end{figure}
%
%%%%%%%%%%%%%%%%%%%%%%

Let us first illustrate the high sensitivity of plasmon parameters to the medium optical dispersion. In Fig.~\ref{fig1}, we plot the frequency-dependent  decay rate $\gamma_{m}(\omega)$ and the radiation efficiency $\eta_{m}(\omega)$ calculated for Au nanorod lengths $2a=20$ nm and $2a=10$ nm. For such parameters, the plasmon radiative decay rate $\gamma_{m}^{rad}(\omega)$, given by Eq.~(\ref{mode-decay-rad-main}), is much smaller than its nonradiative decay rate $\gamma_{m}(\omega)=2\varepsilon''(\omega)/[\partial\varepsilon'(\omega_{m})/\partial \omega_{m}]$, which does not depend on plasmonic system's size or shape. In this case, the full plasmon decay rate, shown in  Fig.~\ref{fig1}(a), can be approximated by $\gamma_{m}(\omega)$ and, hence, the plasmon radiative efficiency, shown in  Fig.~\ref{fig1}(b), is $\eta_{m}(\omega)\approx \gamma_{m}^{rad}(\omega)/\gamma_{m}(\omega)$. Note that $\gamma_{m}(\omega)$ and $\gamma_{m}^{rad}(\omega)$ depend on $\varepsilon''(\omega)$ and $\varepsilon'(\omega)$, respectively, while $\gamma_{m}^{rad}(\omega)$ additionally includes the standard frequency factor $\omega^{3}$ [see Eq.~(\ref{mode-decay-rad-main})]. For comparison, we also show the  Markov approximation results obtained by setting  $\varepsilon(\omega)\rightarrow \varepsilon(\omega_{m})$ while keeping intact the above $\omega^{3}$-dependence in $\gamma_{m}^{rad}(\omega)$.

It is clearly seen in Figs.~\ref{fig1}(a) and \ref{fig1}(b) that both plasmon's decay rate and radiation efficiency deviate from their corresponding Markov approximation values. The dip in the frequency dependence of $\gamma_{m}(\omega)$, seen in  Fig.~\ref{fig1}(a), can be traced to the minimum of $\varepsilon''(\omega)$ for Au  at the plasmon wavelength $\lambda_{m}\approx 675$ nm \cite{johnson-christy}, as discussed above. However, for the plasmon decay rate, the difference between $\gamma(\omega)$ and its Markov approximation value $\gamma_{m}=\gamma(\omega_{m})$, shown by the dashed line in Fig.~\ref{fig1}(a),  is not very significant within the plasmon bandwidth $\approx 4.5\times 10^{-2}\omega_{m}$. In contrast, the radiation efficiency $\eta_{m}(\omega)$ changes substantially (by $\approx 30 \%$) in the same frequency interval [see Fig.~\ref{fig1}(b)]. Importantly,  here $\eta_{m}(\omega)$ exhibits the \textit{opposite} behavior, as compared with the Markov approximation result: it \textit{decreases} as the frequency $\omega$ passes through the plasmon resonance $\omega_{m}$. This change of behavior is due to frequency dependence of the metal susceptibility in plasmonic frequency range, $\chi'(\omega)\propto \omega^{-2}$, leading to $\mu_{m}^{2}(\omega)\propto \omega^{-4}$ and, hence, $\gamma_{m}^{rad}(\omega)\propto \mu_{m}^{2}(\omega)\omega^{3}\propto \omega^{-1}$. In fact, the precise behavior of $\eta_{m}(\omega)$ is more complicated due to nonmonotonic frequency dependence of $\gamma_{m}(\omega)$. Note that  radiation efficiency for the longer nanorod ($2a =20$ nm) is significantly higher than for  the shorter one ($2a =10$ nm) due to larger plasmon dipole moment in the former. As we show below, the effect of metal's optical dispersion on  plasmon radiation efficiency leads to dramatic changes in the shape of emission spectra for  hybrid plasmonic systems in the strong coupling regime.

In Fig.~\ref{fig2}, we include a QE situated at a distance $d$ from the nanorod tip with its dipole moment oriented along the nanorod symmetry axis and  plot  the distance dependence of normalized coupling parameter $g_{em}$ calculated using Eq.~(\ref{coupling-mode-volume-main}). The overall magnitude of QE-plasmon coupling is significantly larger for shorter ($2a=10$ nm) nanorod due to a much stronger field confinement   near smaller nanostructures \cite{shahbazyan-prb18}. With decreasing $d$, as QE approaches  hot spot near the tip, the coupling $g_{em}$ reaches values  $\approx\gamma_{m}/4$, implying that the hybrid system transitions to strong coupling regime \cite{shahbazyan-nl19}. Note that although, for the smaller nanorod, the coupling $g_{em}$ is considerably stronger, its radiation efficiency is significantly lower [see Fig.~\ref{fig1}(a)]; together, these effects give rise to the interference between QE and plasmon optical dipoles which  affects the shape of emission spectra, as we discuss below.

In Figs.~\ref{fig3} and \ref{fig4}, we show the  emission spectra for a QE near the Au nanorod of overall length $2a=20$ nm. In Fig. \ref{fig3}(a), we present the results obtained in the Markov approximation, i.e., by setting $\varepsilon(\omega)\rightarrow \varepsilon(\omega_{m})$ and, accordingly, $\gamma_{m}= \gamma_{m}(\omega_{m})$ and  $\mu_{m}= \mu_{m}(\omega_{m})$ in Eq.~(\ref{spectrum}). For comparison, in Fig.~\ref{fig3}(b), we show same emission spectra but for QE \textit{uncoupled} from the radiation field (i.e., for $\mu_{e}=0$), which is equivalent to the CO model \cite{pelton-oe10,pelton-nc18} albeit with microscopic coupling parameters. As the QE approaches hot spot near the  tip, the system undergoes transition to strong coupling regime signaled  by the emergence of Rabi splitting between polaritonic bands. In both cases, the emission spectra show a distinct \textit{asymmetry} pattern as the upper (higher frequency) polaritonic band is enhanced relative to the lower one. Such an asymmetry is caused by the standard prefactor $\omega^{4}$ in the emission spectrum (\ref{spectrum}) which reflects a faster emission rate for higher energy transitions. The asymmetry is strongest in the CO model approximation [see Fig.~\ref{fig3}(b)], i.e., if only the plasmon's dipole radiates; since, in the Markov approximation, the plasmon's dipole moment stays unchanged across  its bandwidth,  the mixed state with higher energy radiates with a faster rate. In Fig.~\ref{fig3}(a), the asymmetry is somewhat reduced due to the interference effects between QE's and plasmon's optical dipoles. However, for a not very small nanorod, the plasmon-induced QE dipole moment is still much smaller than the plasmon dipole moment, so that, in the Markov approximation, the upper polaritonic band carries a larger spectral weight.

In Fig.~\ref{fig4}(a), we show the results of full calculations incorporating frequency-dependent metal dielectric function carried for the same sets of parameters as in Fig~\ref{fig3}. For comparison, in Fig. \ref{fig4}(b), we show  only the plasmonic antenna contribution obtained by setting $\mu_{e}=0$ in Eq.~(\ref{spectrum}). The striking difference between the emission spectra shown in Fig.~\ref{fig4} and  in Fig.~\ref{fig3} is the \textit{inversion} of  asymmetry pattern as the main spectral weight now rests with the \textit{lower} polaritonic band. Such a shift of  spectral weight can be traced to the effect of metal dispersion on plasmon's optical dipole moment $\mu_{m}(\omega)$ which results in the suppression of plasmon radiation efficiency for higher frequencies, in contrast with the Markov approximation result [see Fig.~\ref{fig1}(b)]. While the asymmetry inversion is clearly visible in the plasmonic antenna spectra [see Fig.~\ref{fig4}(b)], it is even more pronounced in the full spectra [see Fig.~\ref{fig4}(a)] because the interference effects between the QE and plasmon optical dipoles further shift the spectral weight towards lower polaritonic band.

In Figs.~\ref{fig5} and \ref{fig6}, we show the results of our calculations for smaller  nanorod ($2a=10$ nm)  characterized by a stronger field confinement near the tip and, accordingly, a  stronger QE-plasmon coupling [see Fig.~\ref{fig2}]. Apart from a considerably larger Rabi splitting, the main distinction of these emission spectra from those for a larger nanorod is a noticeably weaker asymmetry for spectra calculated in the Markov approximation [compare Fig.~\ref{fig3}(a) and Fig.~\ref{fig5}(a)]. In fact, for intermediate coupling (dashed line), the lower polaritonic band is slightly enhanced [see Fig.~\ref{fig5}(a)]. This can be traced to a combined effect of enhanced plasmon-induced QE dipole moment and reduced plasmon dipole moment, both taking place due to small nanorod size, which gives rise to the interference between plasmon's and plasmon-induced QE's optical dipoles. A similar interference-induced spectral weight shift was recently reported in the weak coupling regime for exciton-induced transparency in scattering spectra \cite{shahbazyan-nanophot21}. 

In Fig.~\ref{fig6}(a) and Fig.~\ref{fig6}(b), we show the emission spectra for  full system and plasmonic antenna,  respectively, calculated with frequency-dependent metal dielectric function. The  asymmetry is more pronounced in  Fig.~\ref{fig6}(a), where  the spectral weight shift is also facilitated by  interference effects. In all cases, the main spectral weight in the emission spectra of a hybrid system firmly rests with the lower polaritonic band, indicating the dominant role of non-Markovian effects in the strong coupling regime.

%Finally, although  non-Markovian effects strongly affect the overall shape of the emission spectra, they also can, in principle, affect the transition point by causing it to occur at a different value of coupling $g_{em}$, especially if the plasmon and QE excitation frequencies are not precisely in resonance. Indeed, for any nanoplasmonic system, the emergence of polatitonic bands is defined by the equation $(\omega-\omega_{e}+\frac{i}{2}\gamma_{e})[\omega-\omega_{m}+\frac{i}{2}\gamma_{m}(\omega)]-g_{em}^{2}=0$ [see Eq.~(\ref{spectrum})], as the real part of its solution splits into two branches at some critical coupling $g_{em}$. Obviously, for $\omega_{m}\neq \omega_{e}$, the critical value of $g_{em}$ can differ from its Markov approximation value if the dispersion of $\gamma_{m}(\omega)$ is sufficiently large [see Fig.~\ref{fig1}(a)]. This effect, however, is expected to be weak and is not considered here.

\section{Conclusions}
\label{sec:conc}

We have studied non-Markovian effects in hybrid plasmonic systems which emerge from optical dispersion of the host material's complex dielectric function. We have found that the emission spectrum of a quantum emitter resonantly coupled to a surface plasmon in a metal-dielectric structure develops a distinct asymmetry pattern as the system transitions to strong coupling regime. By using a novel quantum approach to interacting plasmons that retains the effects of optical dispersion in the coupling parameters \cite{shahbazyan-prb21}, we derived an analytical expression for the system emission spectrum for a plasmonic system of arbitrary shape with characteristic size below the diffraction limit. We analyzed both weak coupling and strong coupling regimes to elucidate the processes involved in the emission of a photon by an excited emitter placed near a plasmonic structure. We found that, in the weak coupling regime, the non-Markovian effects are weak and do not significantly affect spectral shape of the emission peak. In contrast, in the strong coupling regime, non-Markovian effects dramatically affect the shape of emission spectra by causing an inversion of spectral asymmetry, as compared with that predicted in classical and quantum models based on the Markov approximation, by shifting  main spectral weight toward the lower polaritonic band.

%%%%%%%%%%%%%%%%%%%%%%%%%%%%%%%
\acknowledgments
This work was supported in part by National Science Foundation  Grants  No. DMR-2000170, No. DMR-1856515,  and No.  DMR-1826886.

 %%%%%%%%%%%%%%%%%%%%%%%%%%%%%%%%%%%%%%%%%%%%%%%%%%% 
 \appendix*

\section{Quantum approach to interacting plasmons in metal-dielectric structures}

\subsection{Classical plasmon modes}
 We consider QEs situated near a plasmonic metal-dielectric structure characterized by a complex dielectric function $\varepsilon (\omega,\bm{r})=\varepsilon' (\omega,\bm{r})+i\varepsilon'' (\omega,\bm{r})$. We assume that the characteristic system size is much smaller than the radiation wavelength, and so the structure supports localized plasmon modes  described by  quasistatic Gauss's equation $\bm{\nabla}\cdot\left [\varepsilon' (\omega_{m},\bm{r})\bm{\nabla} \Phi_{m}(\bm{r})\right ]=0$, where $\Phi_{m}(\bm{r})$ is the mode potential, chosen to be real here, which defines the mode electric fields as  $\bm{E}_{m}(\bm{r})=-\bm{\nabla} \Phi_{m}(\bm{r})$,  and $\omega_{m}$ is the plasmon mode frequency. In such structures, the plasmon lifetime is mainly determined by the Ohmic losses in metal  incorporated in the classical plasmon Green function tensor \cite{shahbazyan-prl16,shahbazyan-prb18}
\begin{align}
\label{dyadic-plasmon}
\bm{D}_{\rm pl}(\omega;\bm{r},\bm{r}') 
=
\sum_{m}\frac{\omega_{m}}{4 U_{m}}\frac{\bm{E}_{m}(\bm{r}) \bm{E}_{m}  (\bm{r}')}{\omega_{m}-\omega -\frac{i}{2}\gamma_{m}(\omega)},
\end{align}
where $U_{m}$ is the plasmon mode energy \cite{landau},
\begin{align}
\label{energy-mode}
U_{m}= \frac{1}{16\pi} \!\int \!  dV     \dfrac{\partial[\omega_{m}\varepsilon'(\omega_{m},\bm{r})]}{\partial \omega_{m}}\, \bm{E}_{m}^{2}(\bm{r}),
% \nonumber\\
%&= \frac{1}{16\pi} 
%\!\int \!  dV     
%\dfrac{\partial[\omega_{m}\varepsilon'(\omega_{m},\bm{r})]}{\partial \omega_{m}}
% \bm{E}_{m}^{2}(\bm{r}) 
%%\partial [\omega_{m}\varepsilon'(\omega_{m},\bm{r})]/\partial \omega_{m},
\end{align}
and $\gamma_{m}(\omega)$ is the frequency-dependent decay rate, 
\begin{equation}
\label{mode-decay}
\gamma_{m}(\omega)=\dfrac{2\!\int \!  dV     
\varepsilon''(\omega,\bm{r})\bm{E}_{m}^{2}(\bm{r}) }{\!\int \!  dV     
[\partial\varepsilon'(\omega_{m},\bm{r})/\partial \omega_{m}]
 \bm{E}_{m}^{2}(\bm{r}) }.
\end{equation}
In structures with  single-metal components, the decay rate is \cite{stockman-review} $\gamma_{m}(\omega)=2\varepsilon''(\omega)/[\partial\varepsilon'(\omega_{m})/\partial \omega_{m}]$.

\subsection{Describing quantum plasmons by projected reservoir modes}

To transition from the classical to quantum description of plasmons, we recall that  in the macroscopic electrodynamics approach \cite{welsch-pra98,welsch-p00,philbin-njp10}, the electric field operator is defined as
\begin{equation}
\label{field-dissip}
\hat{\bm{E}}(\bm{r})=\!\int_{0}^{\infty}\! d\omega \! \!\int\! dV'\! \bm{D}(\omega;\bm{r},\bm{r}')\hat{\bm{P}}_{N}(\omega,\bm{r}') + \text{H.c.},
\end{equation}
where $\hat{\bm{P}}_{N}(\omega,\bm{r}) = (i/2\pi)\sqrt{\hbar\varepsilon'' (\omega,\bm{r})}\hat{\bm{f}}(\omega,\bm{r})$ is the reservoir noise polarization vector operator and  $\bm{D}(\omega;\bm{r},\bm{r}')$ is the classical electromagnetic Green function. The  noise operators $\hat{\bm{f}}(\omega,\bm{r})$  are driven by the Hamiltonian 
$\hat{H}_{N}= \!\int_{0}^{\infty}\! d\omega \!\int\! dV \,\hbar \omega \hat{\bm{f}}^{\dagger}(\omega,\bm{r}) \cdot \hat{\bm{f}}(\omega,\bm{r})$ and obey the commutation relations 
$[\hat{\bm{f}}(\omega,\bm{r}),\hat{\bm{f}}^{\dagger}(\omega',\bm{r}')]=\bm{I}\delta(\omega-\omega')\delta(\bm{r}-\bm{r}')$, where $\bm{I}$ is the unit tensor.  Then, the plasmon mode expansion of the electric field operator is obtained by inserting the plasmon Green's function (\ref{dyadic-plasmon}) into Eq.~(\ref{field-dissip}). The result can be presented in the form \cite{shahbazyan-prb21}
\begin{equation}
\label{field-plas}
\hat{\bm{E}}_{\rm pl}(\bm{r})=\sum_{m}\!\int_{0}^{\infty}\! d\omega \lambda_{m}(\omega) \hat{b}_{m}(\omega) \tilde{\bm{E}}_{m}(\bm{r}) + \text{H.c.},
\end{equation}
where $\tilde{\bm{E}}_{m}(\bm{r})=\sqrt{\hbar\omega_{m}/4U_{m}} \bm{E}_{m}(\bm{r})$ is normalized mode field, $\lambda_{m}(\omega)$ is a complex function of the form
\begin{align}
\label{lambda}
\lambda_{m}(\omega)= \sqrt{\frac{\gamma_{m}(\omega)}{2\pi}}\frac{ i }{\omega -\omega_{m}+\frac{i}{2}\gamma_{m}(\omega)},
\end{align}
which has a plasmon pole in the complex-frequency plane, and the operator $\hat{b}_{m}(\omega)$ is defined by  projecting  the reservoir noise operator $\hat{\bm{f}}(\omega,\bm{r})$ on a plasmon mode as
\begin{equation}
\label{b-definition}
\hat{b}_{m}(\omega)
%=\frac{\hat{f}_{m}(\omega)}{\sqrt{\gamma_{m}(\omega)}}
=-\frac{\int\! dV\sqrt{\varepsilon'' (\omega,\bm{r})} \,\bm{E}_{m}(\bm{r}) \! \cdot \! \hat{\bm{f}}(\omega,\bm{r})} {\left [\int\! dV\varepsilon'' (\omega,\bm{r}) \,\bm{E}_{m}^{2}(\bm{r})\right ]^{1/2}}.
\end{equation}
Commutation relations for  projected reservoir mode (PRM) operators $\hat{b}_{m}(\omega)$ follow from those for $\hat{\bm{f}}(\omega,\bm{r})$, 
\begin{equation}
\label{b-comm}
[\hat{b}_{m}(\omega),\hat{b}_{n}^{\dagger}(\omega')]=\delta_{mn}\delta(\omega-\omega'), 
\end{equation}
while their time-evolution is driven by the Hamiltonian
\begin{equation}
\label{H-reduced-noise}
\hat{H}_{\rm b}=\sum_{m}  \!\int_{0}^{\infty}\!   d\omega \,\hbar\omega \,\hat{b}^{\dagger}_{m}(\omega)\hat{b}_{m}(\omega), 
 \end{equation} 
where we used  the  relation $\int \! dV\varepsilon''(\omega,\bm{r})\bm{E}_{m}(\bm{r})\cdot\bm{E}_{n}(\bm{r})=0$ for $m\neq n$, which reflects the absence of dissipation-induced coupling between the plasmon modes \cite{shahbazyan-prb21}. In this way, the excessive degrees of freedom of full reservoir Hilbert space are eliminated and the plasmon dynamics is driven by the Hamiltonian (\ref{H-reduced-noise}) in the reduced Hilbert space spanned by the PRM operators $\hat{b}_{m}(\omega)$.

The above approach can be further related to the canonical quantization scheme by defining the plasmon annihilation operators as \cite{shahbazyan-prb21} $\hat{a}_{m}=\!\int_{0}^{\infty}\! d\omega \lambda_{m}(\omega) \hat{b}_{m}(\omega)$, so that the field operator (\ref{field-plas})  has the normal-mode expansion of the form $\hat{\bm{E}}_{\rm pl}(\bm{r})=\sum_{m}\bigl[\hat{a}_{m}\tilde{\bm{E}}_{m}(\bm{r}) + \text{H.c.}\bigr]$. Furthermore, with help of Eq.~(\ref{b-comm}), the canonical commutation relations  $[\hat{a}_{m}, \hat{a}_{n}^{\dagger}]=\delta_{mn}$ can be proven in the Markov approximation by ignoring the dielectric function dispersion, i.e., by setting $\omega=\omega_{m}$ in $\varepsilon(\omega,\bm{r})$. Using the  normal-mode expansion, the canonical plasmon Hamiltonian $H_{\rm pl}=\sum_{m}\hbar\omega_{m}\hat{a}_{m}^{\dagger}\hat{a}_{m}$ can be derived as well together with the plasmon coupling to QEs and the electromagnetic field \cite{shahbazyan-prb21}. However, the canonical quantization scheme is valid only in the Markov approximation and, hence, is not suitable for describing any effects sensitive to strong optical dispersion and losses  in metals.

\subsection{Interactions with quantum emitters}
 
The  interactions  between plasmons and QEs, which are described by two-level systems  with the excitation frequency $\omega_{e}$ positioned at $\bm{r}_{i}$,  are defined by the  Hamiltonian  term  $\hat{H}_{\rm pl-qe}=-\sum_{i}\hat{\bm{p}}_{i}  \cdot \hat{\bm{E}}_{\rm pl}(\bm{r}_{i})$, where $\hat{\bm{p}}_{i}= \bm{\mu}_{i}(\hat{\sigma}^{\dagger}_{i}+\hat{\sigma}_{i})$ is the QE dipole moment. Here,  and $\bm{\mu}_{i}=\mu_{e} \bm{n}_{i}$   is the transition matrix element ($\bm{n}_{i}$ is dipole orientation).  Using the mode expansion for plasmon field operator (\ref{field-plas}),  the coupling Hamiltonian in the rotating wave approximation (RWA) takes the form
\begin{equation}
\label{H-b-qe}
\hat{H}_{\rm b-qe}=\sum_{im} \int_{0}^{\infty}\!  d\omega\left [\hbar q_{im}(\omega)\,\hat{\sigma}^{\dagger}_{i}\,\hat{b}_{m}(\omega) +\text{H.c.}
%g_{im}^{*}(\omega)\hat{b}^{\dagger}_{m}(\omega)\hat{\sigma}_{i}
\right ],
\end{equation}
where $q_{im}(\omega)=g_{im}\lambda_{m}(\omega)$ is the QE-PRM coupling. Here,  $\lambda_{m}(\omega)$ is given by Eq.~(\ref{lambda}) and $g_{im}=-\bm{\mu}_{i}\!\cdot\!\tilde{\bm{E}}_{m}(\bm{r}_{i})/\hbar$ is the standard  QE-plasmon coupling. Note that in terms of the original mode fields,  the QE-plasmon coupling can be recast in a cavity-like form
\begin{equation}
\label{coupling-mode-volume}
g^{2}_{im}
%=\frac{4\pi \omega_{m}\mu^{2}[\bm{n}_{i}\!\cdot\!\bm{E}_{m}(\bm{r}_{i})]^{2}}{\hbar\int \! dV [\partial (\omega_{m}\varepsilon')/\partial \omega_{m}]\bm{E}^{2}_{m}}
=\frac{2\pi \mu_{e}^{2}\omega_{m}}{\hbar{\cal V}^{(i)}_{m}},
~~
\frac{1}{{\cal V}^{(i)}_{m}}
%=\frac{[\bm{n}\!\cdot\!\bm{E}(\bm{r})]^{2}}{8\pi U }
= \frac{2[\bm{n}_{i}\!\cdot\!\bm{E}_{m}(\bm{r}_{i})]^{2}}{\int \! dV [\partial (\omega_{m}\varepsilon')/\partial \omega_{m}]\bm{E}_{m}^{2}},
\end{equation}
where ${\cal V}^{(i)}_{m}$ is the projected  plasmon mode volume  \cite{shahbazyan-acsphot17,shahbazyan-prb18}, which characterizes  the plasmon field confinement at a point $\bm{r}_{i}$ in the direction $\bm{n}_{i}$. To establish a connection to classical enhancement effects, we note that the rate of energy transfer (ET) from an excited emitter to the plasmonic system is given by the first-order probability rate $\Gamma_{i} (\omega)=\sum_{m}\gamma_{i\rightarrow m} (\omega)$, where 
\begin{equation}
\label{rate-abs}
\gamma_{i\rightarrow m} (\omega)=\frac{2\pi}{\hbar}
\int_{0}^{\infty}\!  d\omega' 
\left | \hbar q_{im}(\omega')\right |^{2}\delta(\hbar\omega'-\hbar\omega),
\end{equation}
and the  integral runs over the PRM's final states. Evaluating the  integral, we obtain
\begin{equation}
\label{rate-et-mode}
\gamma_{i\rightarrow m} (\omega)
= 2\pi g_{im}^{2}\left | \lambda_{m}(\omega) \right |^{2}=\frac{g_{im}^{2} \gamma_{m}(\omega)}{(\omega_{m}-\omega)^{2} +\frac{1}{4}\gamma_{m}^{2}(\omega)},
\end{equation}
where we used Eq.~(\ref{lambda}). In fact, the above expression represents the ET rate from a QE to plasmons evaluated, in a standard way,   using the classical plasmon Green's function (\ref{dyadic-plasmon}) as $\Gamma_{i} (\omega)
=(2/\hbar)\text{Im} \left [\bm{\mu}_{i} \bm{D}_{\rm pl}(\omega;\bm{r}_{i},\bm{r}_{i})\bm{\mu}_{i} \right ]$, implying that QE-PRM interactions are mediated by classical plasmons absorbing the QE energy. At resonance  $\omega=\omega_{m}$, we recover the relation between the   QE-plasmon ET rate  and  the  QE-plasmon coupling as \cite{shahbazyan-nl19} $\gamma_{i\rightarrow m} =4g_{im}^{2}/\gamma_{m}$.

\subsection{Interactions with the electromagnetic field}

The coupling to the EM field $\bm{E}(t)$ is described by the Hamiltonian term  $H_{\rm int}=-\text{Re} \int dV \hat{\bm{E}}_{\rm pl}(\bm{r})  \cdot  \bm{P}(t,\bm{r})$, where $\bm{P} =\hat{\chi} \bm{E}$ is the induced polarization vector and $ \chi(t,\bm{r})$ is the plasmonic system susceptibility. For a monochromatic field of frequency $\omega$  which is uniform on the system scale,  using the mode expansion  for the electric field operator (\ref{field-plas}), we obtain in the RWA
\begin{equation}
\label{H-b-em}
\hat{H}_{\rm b-em}= -\!\sum_{m}\! \int_{0}^{\infty}\! \! d\omega' \! \left [\bm{d}_{m}^{*}(\omega,\omega')\!\cdot \!\bm{E}e^{-i\omega t} \hat{b}_{m}^{\dagger}(\omega') +\text{H.c.}
\right ],
\end{equation}
where $\bm{d}_{m}(\omega,\omega')=\bm{\mu}_{m}(\omega)\lambda_{m}(\omega')$ is PRM-EM transition matrix element and $\bm{\mu}_{m}(\omega)=\!\int\! dV\chi' (\omega,\bm{r})\tilde{\bm{E}}_{m}(\bm{r})$ is the plasmon optical dipole moment that determines its frequency-dependent radiative decay rate as \cite{shahbazyan-prb18}
\begin{equation}
\label{mode-decay-rad}
\gamma_{m}^{rad}(\omega)=\frac{4\bm{\mu}_{m}^{2}(\omega)\omega^{3}}{3\hbar c^{3}},
\end{equation}
where $c$ is the speed of light. The  rate of EM field energy absorption has the form 
\begin{equation}
\label{rate-abs-mode}
\Gamma_{m}(\omega)=\frac{2\pi}{\hbar}
\!\int_{0}^{\infty}\!  d\omega' 
\left |  \bm{d}_{m}(\omega,\omega')\!\cdot \!\bm{E}\right |^{2}\delta(\hbar\omega'-\hbar\omega),
\end{equation}
which, after evaluating the integral, can be presented as 
\begin{equation}
\label{rate-abs-pol}
\Gamma_{m} (\omega)
%=\dfrac{2\pi}{\hbar^{2}}\left |  \bm{d}_{m}(\omega,\omega)\!\cdot \!\bm{E}\right |^{2}
=\dfrac{2}{\hbar}\, \text{Im} \left [\bm{E}^{*}\bm{\alpha}_{m}(\omega)\bm{E} \right ].
\end{equation}
Here, $\bm{\alpha}_{m} (\omega)$ is the optical polarizability tensor of a plasmon mode that defines its response to an external field, 
\begin{equation}
\label{polar-mode}
\bm{\alpha}_{m} (\omega)=\frac{1}{\hbar}\frac{\bm{\mu}_{m}(\omega)\bm{\mu}_{m}(\omega)}{\omega_{m}-\omega -\frac{i}{2}\gamma_{m}(\omega)},
\end{equation}
and the full absorption rate is obtained by summing over all PRM modes  as $\Gamma_{\rm pl} (\omega)
=(2/\hbar)\text{Im} \left [\bm{E}^{*}\bm{\alpha}_{\rm pl}(\omega)\bm{E} \right ]$, where $\bm{\alpha}_{\rm pl}(\omega)=\sum_{m}\bm{\alpha}_{m}(\omega)$ is full optical polarizability of the plasmonic structure \cite{shahbazyan-prb18}. Thus, within this approach,  the classical effect of resonant light absorption by surface plasmons is recovered from the Hamiltonian (\ref{H-b-em}) in the first order. We stress that, in contrast to canonical quantization scheme, here the optical dispersion of dielectric function is controlled by the incident light.

Finally, the system Hamiltonian should be supplemented by the QE Hamiltonian
\begin{equation}
\label{H-qe}
H_{\rm qe}=\hbar\omega_{e}\!\sum_{i}\hat{\sigma}_{i}^{\dagger}\hat{\sigma}_{i}
\end{equation}
and the Hamiltonian term describing  QEs' interactions with the EM field,
\begin{equation}
\label{H-qe-em}
H_{\rm qe-em}=-\sum_{i}(\bm{\mu}_{i}\!\cdot\!\bm{E}e^{-i\omega t} \hat{\sigma}_{i}^{\dagger} + {\rm H.c.}).
\end{equation}
The full Hamiltonian $\hat{H}=\hat{H}_{\rm b}+\hat{H}_{\rm b-qe}+\hat{H}_{\rm b-em}+\hat{H}_{\rm qe}+\hat{H}_{\rm qe-em}$ provides a starting point for studying  non-Markovian effects in hybrid plasmonic systems.

%%%%%%%%%%%%%%%%%%%%%%%%%%%%%%%%%%%%%

\end{document}